\newif\ifloguseIEEEConf
\ifloguseIEEEConf
    \documentclass[conference,10pt]{IEEEtran}
\else
    \documentclass{article}
    \usepackage{appendix}
\fi
\usepackage[letterpaper,top=2.6cm,bottom=2.6cm,left=2.25cm, right=2.25cm]{geometry}
\usepackage{graphicx}
\usepackage{amssymb}
\usepackage{pdfsync}
\usepackage{subcaption}

\usepackage{amsfonts,amsmath, amssymb, graphics,geometry,framed,setspace,multirow,slashbox}
\begin{document}

\newtheorem{theorem}{Theorem}
\newtheorem{corollary}{Corollary}
\newtheorem{lemma}{Lemma}
\newtheorem{example}{Example}
\newtheorem{definition}{Definition}
\newtheorem{proposition}{Proposition}
\newtheorem{observation}{Observation}
\newtheorem{conjecture}{Conjecture}
\newtheorem{remark}{Remark}
\newcommand{\N}{\mathbb{N}}
\newcommand{\qed}{\hfill $\diamondsuit$}
\newcommand{\noam}{Noam Presman\ }
\newcommand{\bhat}{Bhattacharyya }
\newcommand{ \etld } {\tilde{\epsilon}}
\newcommand{ \He } {\mathcal{H}}

\ifloguseIEEEConf
 \newcommand{ \proofIdea }{\textit{Proof Idea: }}
\else
  \newcommand{\QED}{\hfill $\diamondsuit$}
  \newcommand{\proof}{\noindent {\bf Proof}\ }
   \newcommand{ \proofIdea }{\textbf{Proof Idea }}
\fi

\title{Polar Codes with Mixed-Kernels } 
\author{Noam Presman, Ofer Shapira and Simon Litsyn\\ School of Electrical Engineering,
Tel Aviv University, Ramat Aviv 69978 Israel. \\e-mails:
 \{presmann, ofershap, litsyn\}@eng.tau.ac.il. }
\date{}

\maketitle

 \begin{abstract}
A generalization of the polar coding scheme called mixed-kernels is introduced.
This generalization exploits several homogeneous kernels over alphabets of different sizes. An asymptotic analysis of the proposed scheme shows that its polarization properties are strongly related to the ones of the constituent kernels. Simulation of finite length instances of the scheme indicate their advantages both in error correction performance and complexity compared to the known polar coding structures.

 \end{abstract}
\section{Introduction}
Polar codes were introduced by Arikan \cite{Arikan} and provide
an error correction scheme for achieving the symmetric capacity of binary
memoryless channels (B-MC) with polynomial encoding and decoding
complexity. Originally, Arikan considered binary and linear polar codes that are based on a two dimensional kernel, known as the $(u+v,v)$ mapping. This mapping is extended  to generate  arbitrary  codes of length $N=2^n$ bits, by a Kronecker power of the generating matrix that defines the transformation. Multiplying a permutation of an $N$ bits input vector $\bf u$ by this matrix results in a vector $\bf x$, that is transmitted over  $N$ independent copies of a memoryless channel, $\mathcal{W}$. As a result, $N$ dependent channels between the components of $\bf u$ and the outputs of the copies of the channel $\mathcal{W}$ are created. These channels exhibit polarization under successive-cancellation (SC) decoding: as $n$ grows there is a proportion of  $I(\mathcal{W})$ (the symmetric channel capacity) of the channels that have their capacity approaching to $1$, while the rest of the channels have their capacity approaching to $0$.

The exponent of the kernel as a measure of the asymptotic rate of polarization for arbitrary binary and linear polar codes was introduced by Korada \textit{et al.}  \cite{Korada2010} and further generalized to arbitrary polarizing kernel by Mori and Tanaka  \cite{Mori2010}. The authors suggested designing binary kernels based on the idea of code decomposition \cite{PrShLi2}. It was noted in that paper  that taking advantage
of the explicit (non-binary) code decomposition in order to construct polar code introduces more flexibility to the design. This however, usually requires utilizing at least two kernels and combining them appropriately. This technique results in a mixed-kernels structure. Our objective in this paper is to explore such mixed-kernels constructions and analyze them.

This paper is organized as follows. In Section \ref{sec:Preliminar} we review the idea of code decomposition and its relation to the
design of polar code kernels. This notion is the main motivation for the
introduction of mixed-kernels.
For simplicity, we decided to first present the concept of  mixed-kernels by an example of a specific construction that is based on a binary kernel and a quaternary
kernel. This is done in Section \ref{sec:mixKer}. The discussion is then broaden in Section \ref{sec:GenAndConc} by introducing general mixed-kernels. Section \ref{sec:mixedPotAdvantage} elaborates on  two advantages that finite length mixed-kernels structures may have over known polar coding schemes: improved code decomposition leading to better error-correction (Subsection \ref{sec:improvCodeDec}) and moderate decoding complexity (Subsection \ref{sec:redDecComplx}).
Simulations results  demonstrating these advantages are given in Section  \ref{sec:simulations}.

Throughout we use the following notations. For a natural number ${\ell}$, we denote
$[{\ell}]=\left\{1,2,3,...,{\ell}\right\}$ and $[{\ell}]_{-}=\left\{0,1,2,...,{\ell}-1\right\}$. We denote vectors in bold letters.  For $i\geq j$, let ${\bf
u}^i_j=\left[u_j \,\,\, u_{j+1} \ldots \,\,\,\, u_i\right]$ be the sub-vector of ${\bf u}$ of length
$i-j+1$ (if $i<j$ we say that ${\bf
u}^i_j=[\,\,\,]$, the empty vector, and its length is $0$). We also occasionally use the following notation $\left[u_k \right]_{k=j}^{k=i}$ to refer to the same sub-vector  ${\bf
u}^i_j$. For two vectors $\bf u$ and $\bf v$ of lengths $n_u$ and $n_v$, we denote the $n_u+n_v$ length vector which is the concatenation of $\bf u$ to $\bf v$ by $[{\bf u} \,\,\,{\bf v}]$ or $[{\bf u},  {\bf v}]$ or just ${\bf u}\bullet{\bf v}$.  For a scalar $x$, the  $n_u+1$  length vector ${\bf u}\bullet x$, is just the concatenation of the vector ${\bf u}$ with the length one vector containing $x$. 

%

\section{Preliminaries}\label{sec:Preliminar}

In this paper we consider kernels that are based on bijective
transformations over a field $F$.
A channel polarization kernel of ${\ell}$ dimensions, denoted by $g(\cdot)$, is a mapping
$$
g(\cdot):F^{{\ell}}\rightarrow F^{{\ell}}.
$$
This means that $g({\bf u})={\bf x}, \,\,\,\, {\bf u}, {\bf x}
\in F^{{\ell}}$.

 We refer to this type of kernel as a \textit{homogeneous kernel}, because its $\ell$ input coordinates and $\ell$ output coordinates are from the same alphabet $F$. Symbols from an alphabet $F$ are called $F$-symbols in this paper. The homogenous kernel $g(\cdot)$ may generate a polar  code of length $\ell^m$ $F$-sybmols by inducing a larger mapping from it, in the following way \cite{Mori2010}.
\begin{definition}[Homogenous Polar Code Generation]\label{def:constructG2}
Given an ${\ell}$ dimensions transformation $g(\cdot)$, we
construct a mapping $g^{(n)}(\cdot)$ of $N={\ell}^n$  dimensions (i.e.
$g^{(n)}(\cdot):F^{{\ell}^n}\rightarrow F^{{\ell}^n}$)
in the following recursive fashion.
$$g^{(1)}({\bf u}_0^{\ell-1})=g({\bf u}_0^{\ell-1})\,\,\,;$$

$$g^{(n)}=\Big[ g\left(\gamma_{0,0}, \gamma_{1,0}, \gamma_{2,0}, \ldots, \gamma_{\ell-1,0}\right),$$
$$\,\,\,\,\,\,\,g\left(\gamma_{0,1}, \gamma_{1,1}, \gamma_{2,1}, \ldots, \gamma_{\ell-1,1}\right),\ldots,$$
$$
\,\,\,\,\,\,\,g\left(\gamma_{0, N/{\ell}-1}, \gamma_{1,  N/{\ell}-1}, \gamma_{2,  N/{\ell}-1}, \ldots, \gamma_{\ell-1, N/{\ell}-1}\right)  \Big],
$$
\normalsize
where
$$
\left[\gamma_{i,j}\right]_{j=0}^{j=N/\ell-1}=g^{(n-1)}\left({\bf u}_{ i \cdot (N/\ell)} ^{(i+1)\cdot (N/\ell) -1}\right),
\,\,\,\,\,\,\,\,  i\in\left[\ell\right]_{-}.
$$
\end{definition}

\subsection{Polar Codes as Recursive General Concatenated Codes}\label{subsect:kernelAndCodeDecompose}
 General Concatenated Codes (GCC)\footnote{The construction of the GCCs is a generalization of  Forney's code concatenation method \cite{Forney1966}.} are error correcting codes that are constructed by a technique, which was introduced by Blokh and Zyabolov \cite{Blokh1974} and Zinoviev \cite{Zinoviev1976}. In this construction, we have $\ell$ outer-codes $\left\{\mathcal{C}_i\right\}_{i=0}^{\ell-1}$, where $\mathcal{C}_i$ is an $N_{out}$ length code of size $M_i$ over alphabet $F_i$. We also have an inner-code of length $N_{in}$ and size $\prod_{i=0}^{\ell-1}|F_i|$ over alphabet $F$, with a nested encoding function $\varphi(\cdot) : F_0\times F_1 \times ... \times F_{\ell-1} \rightarrow F^{N_{in}}$. The GCC that is generated by these components is a code of length  $N_{out}\cdot N_{in}$ $F$-symbols and of size  $\prod_{i=0}^{\ell-1}M_i$. It is created by taking an $\ell\times N_{out}$ matrix, in which the $i^{th}$ row is a codeword from $\mathcal{C}_i$, and applying the inner mapping $\phi$ on each of the $N_{out}$ columns of the matrix. As Dumer describes in his survey \cite{DumerConcatCodes}, the GCCs can provide good code parameters  for short length codes when using a good combination of outer-codes and a nested inner-code. In fact, some of them give the best parameters known. Moreover,  decoding algorithms may utilize their structure by performing local decoding steps  on the (short) outer-codes and utilizing the inner-code layer for exchanging decisions between the outer-codes.

  As Arikan already noted, polar codes are examples of recursive GCCs \cite[Section I.D]{Arikan}. This observation is useful as it allows to formalize the construction of a large length polar code as a concatenation of several smaller length polar codes (outer-codes) by using a kernel mapping (an inner-code). Therefore, applying this notion to Definition \ref{def:constructG2}, we observe that a polar code of length $N=\ell^n$ symbols, may be regarded as a collection of $\ell$ outer polar codes of length $\ell^{n-1}$ (the $i^{th}$ outer-code   is $\left[\gamma_{i,j}\right]_{j=0}^{j=N/\ell-1}=g^{(n-1)}\left({\bf u}_{i\cdot N/{\ell}}^{(i+1)\cdot N/\ell-1}\right)$ for $i\in \left[\ell\right]_{-}$). These codes are then joined together by employing an inner-code (defined by the function $g(\cdot)$) on the outputs of these mappings. There are $N/\ell$ instances of the inner-mapping, such that instance number $j\in \left[N/{\ell}\right]_{-}$ is applied on the   $j^{th}$ symbol  from each outer-code.

\begin{figure}
\center
  \includegraphics[scale = 0.13]{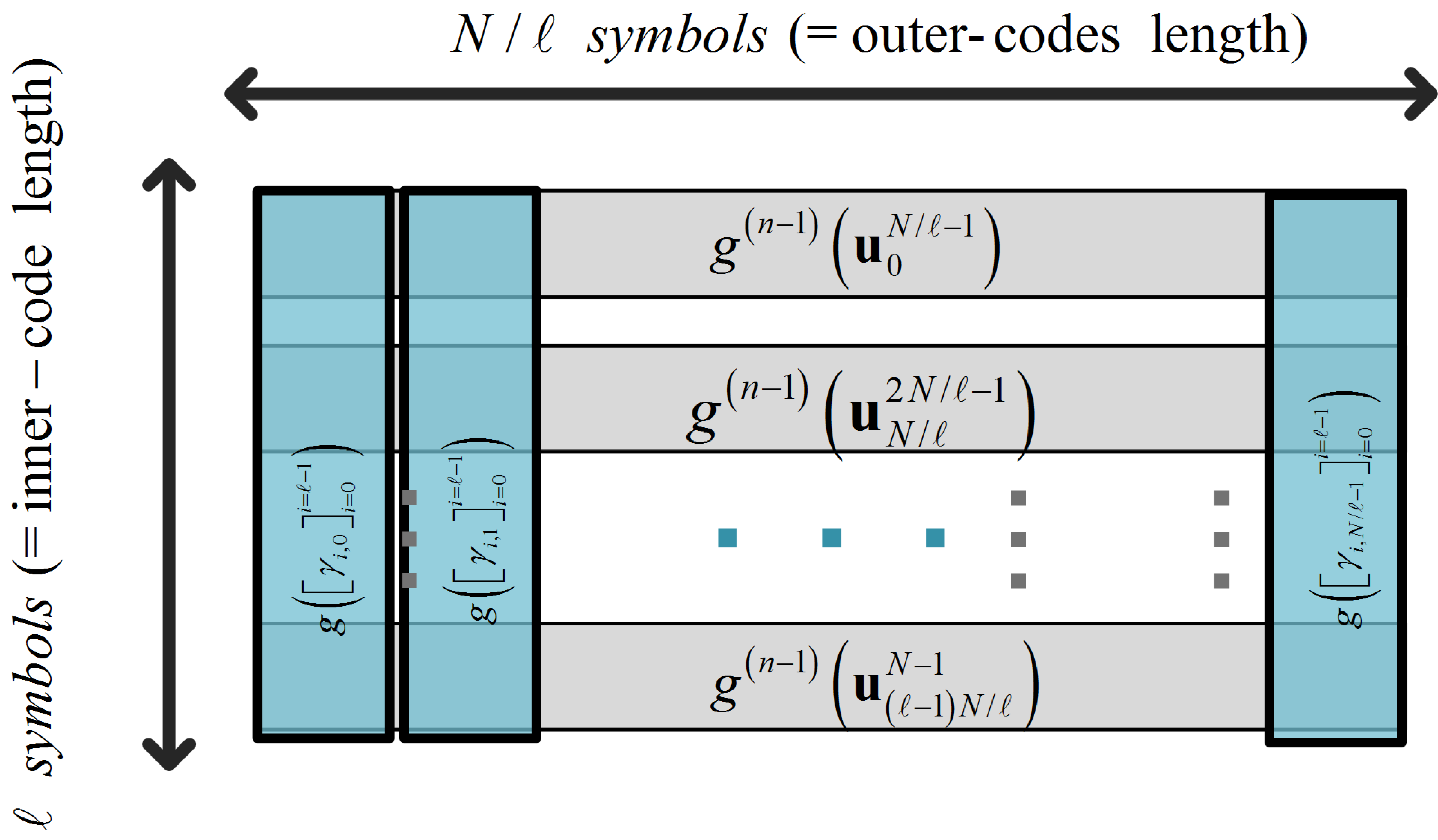}\\
  \caption{GCC representations of a polar code of length $\ell^n$ symbols constructed by a homogenous kernel according to Definition \ref{def:constructG2}   }\label{fig: def2GCC}
\end{figure}

  The above GCC formalization is illustrated in Figure \ref{fig: def2GCC}. In this figure, we see the $\ell$ outer-codewords of length $\ell^{n-1}$ depicted as  gray horizontal rectangles (resembling rows of a matrix). The instances of the inner-codeword mapping are depicted as  vertical rectangles that are located on top of the gray outer-codes rows (resembling columns of a matrix). This is appropriate,  as this mapping operates on columns   of the matrix which rows are the outer-codewords. Note, that for brevity we only drew three instances of the inner mapping, but there should be $\ell^{n-1}$ instances of it, one for each column of this matrix. In the homogenous case, the outer-codes themselves are constructed in the same manner. However, note that even though the outer-codes have the same structure, they are different codes in the general case. The reason is that they may have different sets of frozen symbols associated with them.

\begin{figure}
\center
  \includegraphics[scale = 0.19]{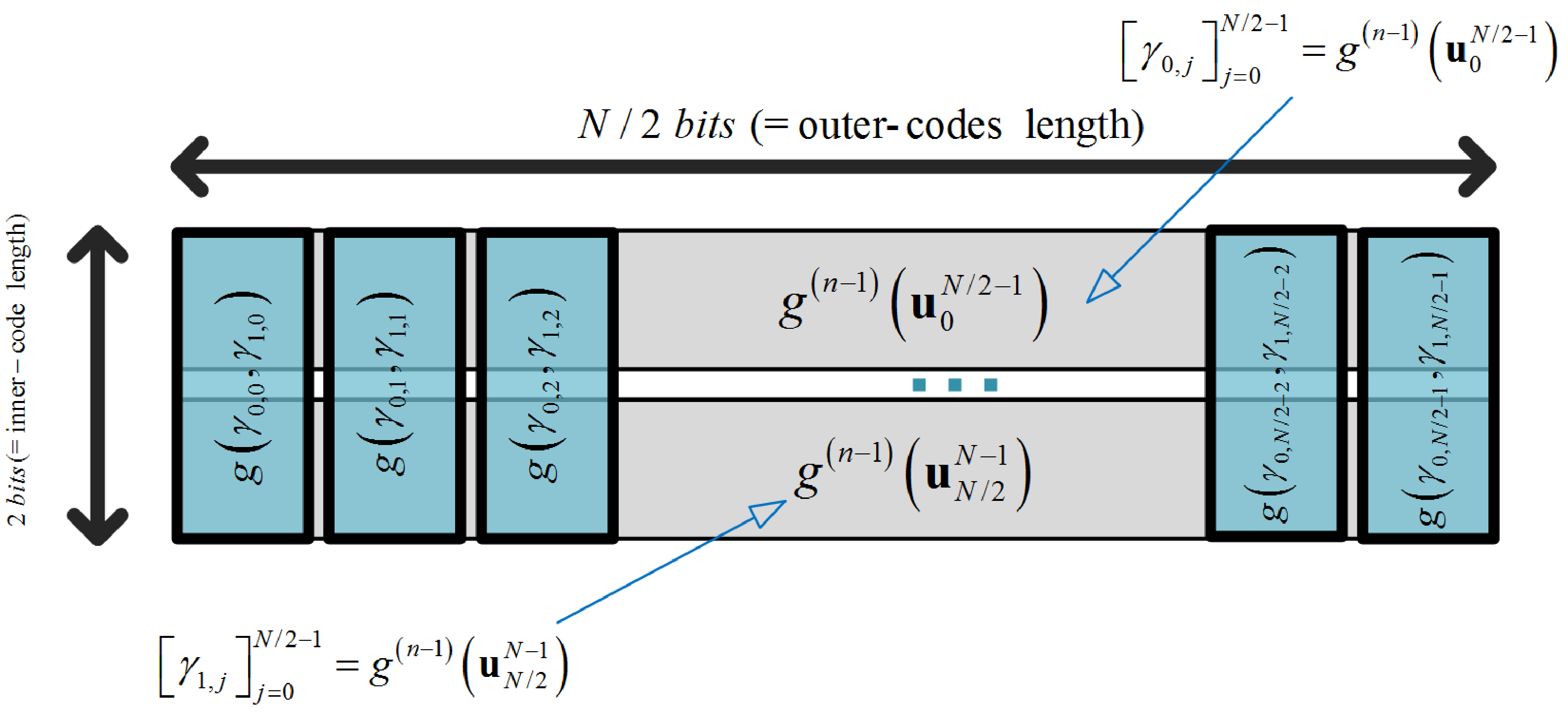}\\
  \caption{GCC representations corresponding to Example \ref{ex:UVV} (Arikan's construction)  }\label{fig:gccUVV}
\end{figure}

\begin{example}[Arikan's Construction]\label{ex:UVV}
Let ${\bf u}$ be an $N=2^n$ length binary vector.  The vector $\bf u$ is transformed into an $N$ length vector $\bf x$ by using a bijective mapping $g(\cdot):\{0,1\}^{N}\rightarrow \{0,1\}^{N}$. The transformation is defined recursively as
$$
\text{for } n=1\,\,\,\,\, g^{(1)}({\bf u})=\left[u_0+u_1,u_1\right]
$$
\begin{equation}\label{eq:Constr}
\text{for }  n>1\,\,\,\,\, g^{(n)}({\bf u})={\bf x}_0^{N-1}\,\,\,\,,
\end{equation}
where $\left[x_{2j}, \,\,\, x_{2j+1}\right] = \left[\gamma_{0,j}+\gamma_{1,j},\,\,\,\,\,\gamma_{1,j}\right]$ for $j\in [N/2]_{-}$, and $\left[\gamma_{0,j} \right]_{j=0}^{N/2-1}=g^{(n-1)}\left({\bf u}_0^{N/2-1}\right)$, $\left[\gamma_{1,j} \right]_{j=0}^{N/2-1}=g^{(n-1)}\left({\bf u}_{N/2}^{N-1}\right)$ are the two outer-codes (each one of length $N/2$ bits). Figure \ref{fig:gccUVV} depicts the GCC block diagram for this example.
\end{example}

The GCC structure of polar codes can be also  represented by a layered\footnote{The vertices of a layered graph can be partitioned into a sequence of sub-sets called layers and denoted by $L_0,L_1,\cdots,L_{k-1}$. The edges of the graph connect only vertices within the same layer or vertices in layers with successive ordinals.} Forney's normal factor graph \cite{Forney2001}.  Layer $\# 0$ of this graph contains the inner mappings (represented as sets of vertices), and therefore we refer to it as the \textit{inner-layer}. Layer $\#1$ contains the vertices of the inner layers of all the outer-codes that are concatenated by layer $\#0$. We may continue and generate layer $\#i$ by considering the outer-codes that are concatenated by layer $\#(i-1)$ and include in this layer all the vertices describing their inner mappings. This recursive construction process may continue until we reach to outer-codes that cannot be decomposed to non-trivial inner-codes and outer-codes.  Edges (representing variables) connect between outputs of the outer-codes to the inputs of the inner mappings. This presentation can be viewed as observing the GCC structure in Figure \ref{fig: def2GCC} from its side.

\begin{figure}
\center
  \includegraphics[scale = 0.20]{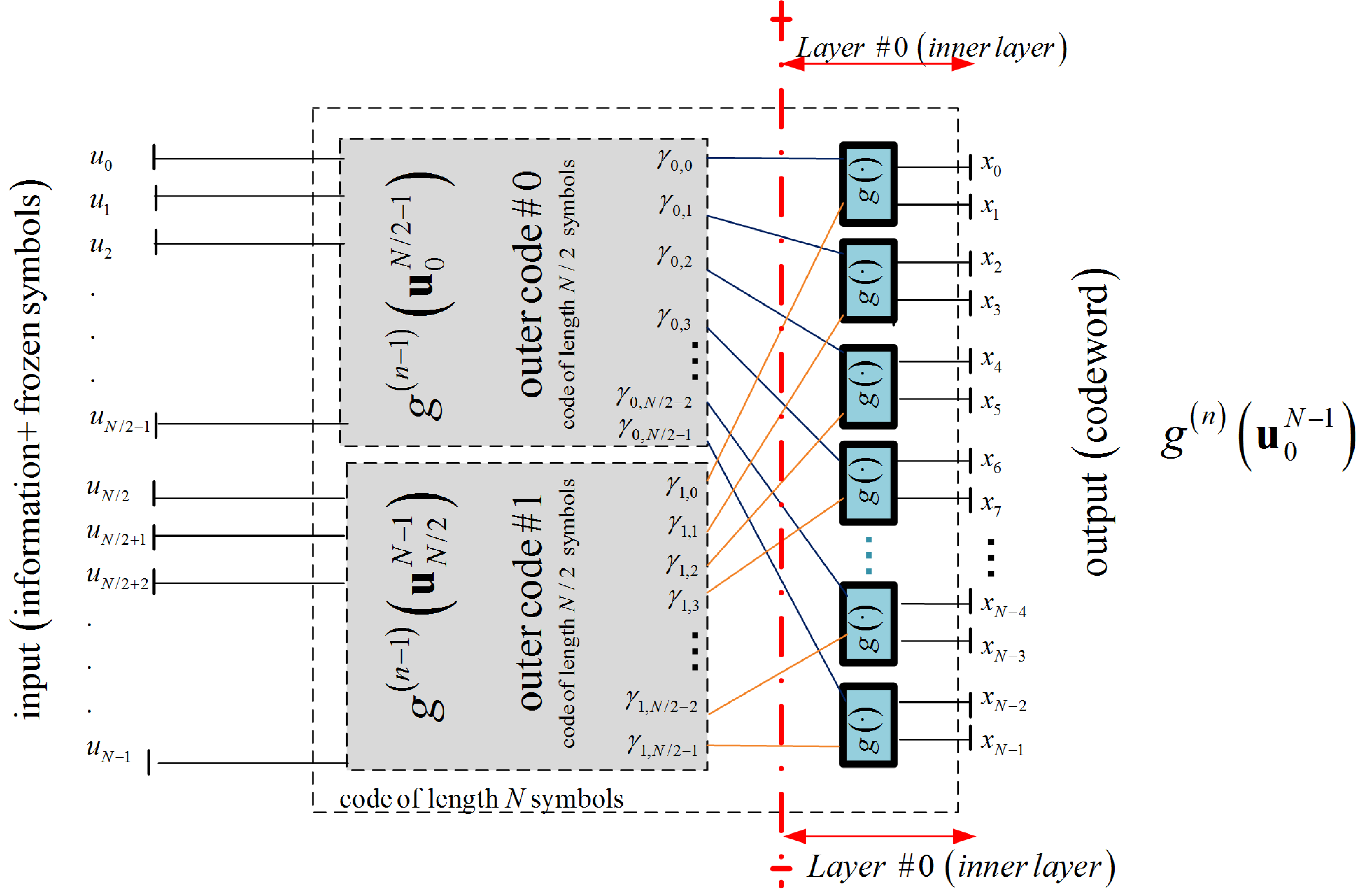}\\
  \caption{ Representation of a polar code with kernel size $\ell = 2$ symbols as a layered factor graph    }\label{fig:GCCWithLayer0}
\end{figure}

\begin{figure}
\center
  \includegraphics[scale = 0.20]{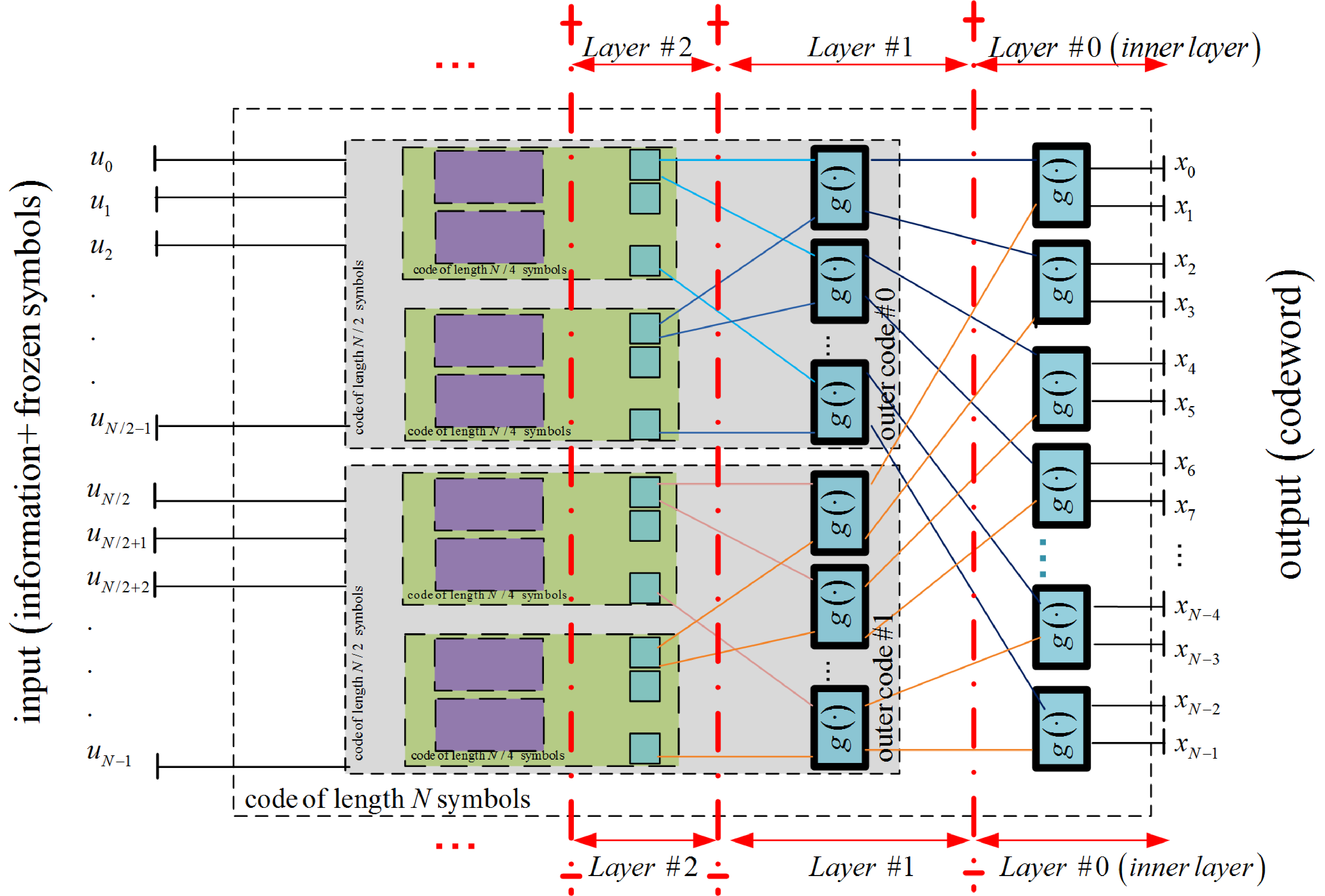}\\
  \caption{ Representation of a polar code with kernel size $\ell = 2$ symbols as a layered factor graph (detailed version of Figure \ref{fig:GCCWithLayer0} - recursion expanded)    }\label{fig:GCCWithLayer}
\end{figure}

\begin{figure}
\center
  \includegraphics[scale = 0.16]{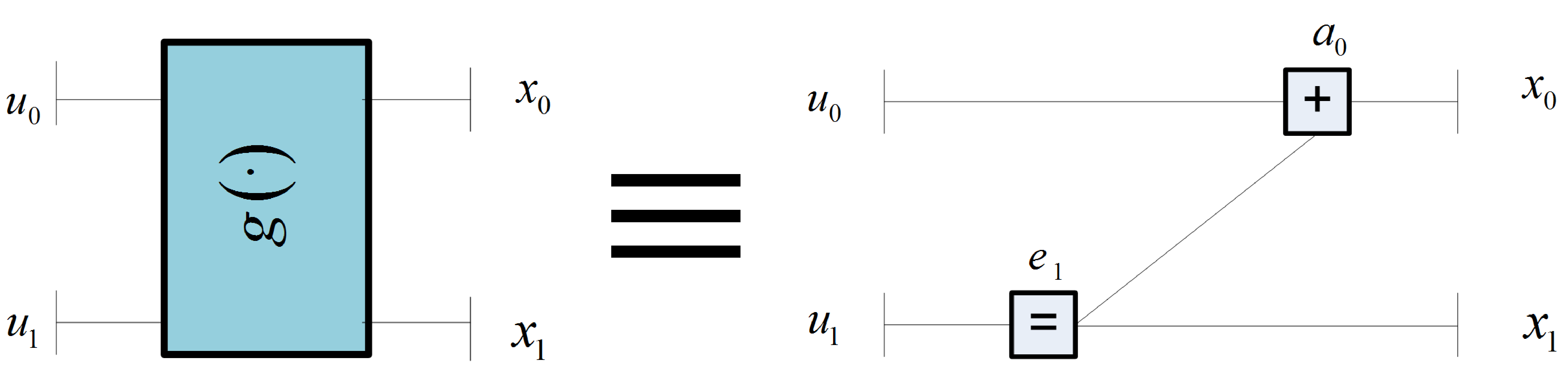}\\
  \caption{ Normal factor graph representation of the $g(\cdot)$ block from Figures \ref{fig:GCCWithLayer0} and \ref{fig:GCCWithLayer} for Arikan's $(u+v,v)$ construction   }\label{fig:GCCWithLayerUVVPart}
\end{figure}

\begin{example}[Layered Normal Factor Graph for Arikan's Construction]\label{ex:ArikanLayeredFactorGraph}
Figures   \ref{fig:GCCWithLayer0} and \ref{fig:GCCWithLayer} depict a layered factor graph representation for length $N=2^n$ symbols polar code with kernel of $\ell = 2$ dimensions. Figure \ref{fig:GCCWithLayer0} gives only a block structure of the graph, in which we have the two outer-codes of length $N/2$ that are connected by the inner layer (note the similarities to the GCC block diagram in Figure \ref{fig:gccUVV}). Half edges represent the inputs ${\bf u}_{0}^{N-1}$ and the outputs  ${\bf x}_{0}^{N-1}$ of the transformation. The edges (denoted by $\gamma_{i,j},\,\,\,\,  j\in\left[N/2\right]_{-} i\in[2]_{-}$) connect the outputs of the two outer-codes to the inputs of the inner mapping blocks,  $g(\cdot)$. A more elaborated version of this figure is given in Figure \ref{fig:GCCWithLayer}, in which we expanded the recursive construction.

Strictly speaking, the green blocks that represent the $g(\cdot)$ inner-mapping are themselves factor graphs (i.e.  collections of vertices and edges). An example of a normal factor graph specifying such a block is given in Figure \ref{fig:GCCWithLayerUVVPart} for  Arikan's $(u+v,v)$ construction (see Example \ref{ex:UVV}). Vertex $a_0$ represents a parity constraint and vertex $e_1$ represents an equivalence constraint.  The half edges $u_0,u_1$ represent the inputs of the mapping, and the half edges $x_0,x_1$ represent its outputs. This graphical structures is perhaps the most popular visual representation of polar codes (see e.g. \cite[Figure 12]{Arikan} and \cite[Figure 5.2]{KoradaThesis} ) and is also known as the "butterflies" graph because of the edges arrangement  in Figure \ref{fig:GCCWithLayer}.
\end{example}


When using SC types of decoders, we sequentially decide on the outer-codes (first on outer-code $\#0$, then on outer-code $\#1$....). The codeword space that is spread by the inner-code mapping is expurgated according to the previous decisions that we made. The latter refined space is used to support decoding of the next outer-code (by calculating the appropriate updated log-likelihood input to its decoder). We say that the inner-code decomposes the codeword space into sub-codes (or partitions). Therefore the SC decoder can be regarded as a sequential decision algorithm on the identity of the partitions to which the transmitted codeword belongs. Therefore, it comes as no surprise that the code decomposition induced by the kernel and its properties play a significant role in the performance of the SC decoder of polar codes.
\subsection{Kernels and Code Decompositions}\label{subsect:kernelAndCodeDecompose}
The notion of code decomposition and its relation to the construction of kernels for polar codes  were previously explored by the authors \cite{PrShLi2}.
We review these concepts here in order to further develop them in the next section.

\begin{definition}[Code Decomposition] \label{def: compoIntoCosets}
Denote by  $T_0^{()} = F^{\ell}$. We perform a sequential partitioning of this set. First, $T_0^{()}$ is partitioned into $f_0\triangleq|F|^{\eta_0}$ subsets, each one of size $\frac{|F|^{\ell}}{f_0}$ denoted by $T_1^{(b_0)}$ where $b_0\in \left[f_0\right]_{-}$. In the next step each one of the sets  $T_1^{(b_0)}$  is decomposed into $f_1\triangleq|F|^{\eta_1}$ subsets, each one of size $\frac{|F|^{\ell}}{f_0\cdot f_1}$. We denote these subsets by $T_1^{([b_0 \,\,\, b_1])}$ where $b_1\in \left[f_1\right]_{-}$. In step $i\in [m]_{-}$,  $T_i^{({\bf b}_0^{i-1})}$  is partitioned into $f_i\triangleq|F|^{\eta_i}$ equally sized subsets $\left\{T_{i+1}^{({\bf b}_0^{i-1}\bullet b_i)}\right\}_{b_i \in [f_i]_{-}}$, of size $\frac{|F|^{{\ell}}}{\prod_{j=0}^{i}f_j}$ ($i\in[m]_{-}$).
 We denote the set of subsets (or sub-codes) of level number $i$ by $T_i$, that is $T_i = \left\{ T_i^{({\bf b}_0^{i-1})}|b_j\in \left[f_j\right]_{-} ,\,\,\,j\in [i]_{-}\right\}$.

 The set $\left\{T_0,...,T_{m}\right\}$ is called a code decomposition of $F^{\ell}$.
The decomposition is commonly  described by the following chain of code parameters $$(\ell,k_0,d_0)-(\ell,k_1,d_1)-...-(\ell,k_{m},d_{m}),$$ if for each $\mathcal{T}\in T_i$ we have that  $\mathcal{T}$ is a code of length $\ell$, size $|F|^{k_i}$ and minimum distance at least $d_i$, for all $i\in [m]_{-}$.

 If the sub-codes of the decompositions are cosets, then we say that $\left\{T_0,...,T_{m}\right\}$ is a decomposition into cosets. In this case, for each $T_i$ the sub-code that contains the zero codeword is called the representative sub-code, and a minimal weight codeword for each coset  is called a coset leader. If all the sub-codes in the decomposition are cosets of linear codes, we say that the decomposition is linear.
\end{definition}

A transformation $g(\cdot)$ can be associated to a code decomposition in the following way.
\begin{definition}[Kernel Definition from Code Decomposition]\label{def: transAssiciatedTheCodes}
Let  $\left\{T_0,...,T_{m}\right\}$ be a code decomposition of $F^{\ell}$ as described in Definition \ref{def: compoIntoCosets}  and such that $\forall \mathcal{T}\in T_{m}, \,\,\,\, |\mathcal{T}|=1$ (i.e. the final step of the decomposition is into singletons).
The  transformation  that is induced by this  code decomposition is defined as follows.
\begin{equation}\label{eq:gMixedDefined}
g(v_0,v_1,...,v_{m-1}):\left(\prod_{i=0}^{m-1}F^{\eta_i}\right)\rightarrow F^{\ell}\,\,\,\,\,\,\,\,\,; \sum_{i=0}^{m-1}\eta_i=\ell\,\,\,\,
\end{equation}
\begin{equation}\label{eq:gBinDefined}
\forall {\bf v} \in \prod_{i=0}^{m-1} F^{\eta_i}, \,\,\,\,\,\,\,g({\bf v}_0^{m-1})={\bf x}_0^{\ell-1} \,\,\,\,\, \text{iff  }\,\,\,\, {\bf
x}_0^{\ell-1} \in T_{m}^{\left({\bf v}_0^{m-1}\right)},
\end{equation}
where in the notation of $T_{m}^{\left({\bf v}_0^{m-1}\right)}$ we take the decimal representation of the components of $\bf v$, for consistency with Definition \ref{def: compoIntoCosets}. In some cases, it is useful to denote the argument of $g(\cdot)$ as  a vector ${\bf u}\in F^{\ell}$, i.e. write $g({\bf u})$ instead of $g({\bf v})$ where ${\bf v}\in\prod_{i=0}^{m-1}F^{\eta_i}$. In this case,
there exists the obvious correspondence between ${\bf v}$ and $\bf u$: $v_i={\bf u}_{s}^{f}$, where $s=\sum_{j=0}^{i-1}\eta_j$, $f=\sum_{j=0}^{i}\eta_j$ and $i\in [m]_{-}$.  We say that $v_i$ is representing $\eta_i$ symbols that are "glued" together. It is convenient to denote $v_i$ as $u_{(s,f)}$, if $v_i={\bf u}_s^f$.
\end{definition}
\begin{example}[Decomposition that Defines  a Kernel]\label{ex:compoAndPolarCode}
In our previous correspondence  \cite[Example 1]{PrShLi2} we considered the decomposition into cosets described by the chain $(4,4,1)-(4,3,2)-(4,1,4)$. Using Definition \ref{def: transAssiciatedTheCodes}, we introduce a kernel function
\begin{equation}\label{eq:g1Defined}
 g_0(u_0,u_{(1,2)},u_3):\left\{0,1\right\}\times\left\{0,1\right\}^2\times\left\{0,1\right\}\rightarrow\left\{0,1\right\}^4 \end{equation}
 that is induced by this decomposition.
The first bit $u_0 $ chooses between  sub-codes $T_1^{(0)}$ and $T_1^{(1)}$. The second and the third bits are glued together, forming
a binary pair, or a quaternary symbol $u_{(1,2)}$ and they indicate the selected sub-code of $T_1^{(u_0)}$. Finally, $u_3$ selects
the codeword from the chosen sub-code. Note  that a straight-forward implementation of the encoding function is to multiply $\bf u$ by the appropriate generating matrix.
\end{example}
There seems to be nothing substantially new in Example \ref{ex:compoAndPolarCode}.  Nevertheless, the challenge here is to
extend this mapping to an $N=4^n$ bits length code. The standard Arikan's  construction (based on the Kronecker power) does not suffice,
because of the glued bits $u_{(1,2)}$, that need to be jointly treated as a quaternary symbol. To facilitate this, we
suggest introducing a second  quaternary kernel, $g_1(\cdot)$. Because different coordinates of the input of $g_0(\cdot)$ are from different alphabet sizes, and because in order to implement this polarization scheme, we incorporate two mapping functions $g_0(\cdot)$ and $g_1(\cdot)$, we refer to the overall construction as a \textit{mixed-kernels} construction.
 Details on how to combine kernels $g_0(\cdot)$ and $g_1(\cdot)$ into a mixed-kernels construction are given in Section \ref{sec:mixKer}. The general construction is presented in Section \ref{sec:GenAndConc}.

\section{Mixed-Kernels by an Example}\label{sec:mixKer}
In this section we introduce the concept of polar codes based on mixed-kernels. In order to have a more comprehensible presentation of the idea we chose to firstly describe specific members of the mixed-kernels ensemble.
This specific example of mixed-kernels seems to be attractive because of its relative simplicity and good error-correction performance as we further observe in Section \ref{sec:simulations}.
The general structure of mixed-kernels may be easily derived from this example and is further discussed in Section \ref{sec:GenAndConc}.
\subsection{Construction of a Mixed-Kernels Polar-Code}\label{sec:mixKerConstruct}

Let $g_0(\cdot)$ be the mapping defined in (\ref{eq:g1Defined}). Let $g_1(\cdot):\left(\{0,1\}^2 \right)^4\rightarrow \left(\{0,1\}^2 \right)^4$ be a polarizing kernel over the quaternary alphabet. For example, $g_1(\cdot)$ can be a kernel, based on the extended Reed-Solomon code of length $4$, $G_{RS}(4)$ that was proven by Mori and Tanaka \cite[Example 20]{MoriandTanka} to be a polarizing kernel (we refer to the code generated by $G_{RS}(4)$ as the $RS4$ polar code). Using $g_1(\cdot)$, we can extend the mapping  of $g_0(\cdot)$ to a length $N=4^n$  bits code. Both $g_0(\cdot)$ and $g_1(\cdot)$ are referred to as the \textit{constituent} kernels of the construction. Note that $g_1(\cdot)$ is introduced in order to handle the glued bits $u_{(1,2)}$ of the input of $g_0(\cdot)$ and therefore is also referred to as the \textit{auxiliary kernel} of the construction.

Let us first review the channel splitting principle \cite[Section I.B]{Arikan} using $g_0(\cdot)$. The output of $g_0(\cdot)$ is binary. We also assume that the channel on which the result  of the transformation (i.e. the codeword) is sent on is binary input and memoryless. The meaning of taking two inputs and glue them together is    that these inputs are treated as a unified entity for decoding and decision making.

Let us denote by $\bf u$  and $\bf x$ two binary vectors that are, respectively, the input and the output of the mapping $g_0(\cdot)$.
$$
g_0(u_0,u_{(1,2)},u_3)={\bf x}_0^3,\,\,\,\,\,\, u_0,u_3\in \{0,1\} ,
$$
$$
u_{(1,2)}\in \{0,1\}^2,x_i\in \{0,1\}\,,i\in[4]_{-}
$$
${\bf x}_0^3$ is transmitted over $4$ copies of the binary memoryless channel $\mathcal{W}$ and  the channel output vector $\bf y$ is received. The channel splitting principle dictates the following synthetic channels and their corresponding transition functions.
\begin{description}
  \item[channel $\mathcal{W}_4^{(0)}$:] $
W_4^{(0)}({\bf y}_0^3|u_0)\triangleq\sum_{u_{(1,2)}\in \{0,1\}^2,u_3\in \{0,1\}}\frac{1}{2^3}\cdot W_4({\bf y}_0^3|u_0,u_{(1,2)},u_3),\,\,\, u_0\in\{0,1\}$.
  \item[channel $\mathcal{W}_4^{(1,2)}$: ] $W_4^{(1,2)}({\bf y}_0^3,u_0|u_{(1,2)})\triangleq\sum_{u_3\in  \{0,1\}}\frac{1}{2^2}\cdot W_4({\bf y}_0^3|u_0,u_{(1,2)},u_3),\,\,\, u_{(1,2)}\in\{0,1\}^2$.
  \item[channel $\mathcal{W}_4^{(3)}$:] $W_4^{(3)}({\bf y}_0^3,u_0,u_{(1,2)}|u_3)\triangleq\frac{1}{2^3} \cdot W_4({\bf y}_0^3|u_0,u_{(1,2)},u_3),\,\,\, u_{3}\in\{0,1\}.$
  \end{description}
Here we use Arikan's notations, according to which  $W_4\left({\bf y}_0^3 |{\bf u}_0^3\right) = \prod_{i=0}^3 W\left(y_i|x_i\right)$, where ${\bf x} = g_0\left({\bf u}\right)$ and $W(y|x)$ is the transition function of the $\mathcal{W}$ channel.

Next, consider  $g_1(\cdot)$, which is a quaternary input and output mapping. A binary vector ${\bf u} \in \{0,1\}^8$ is transformed into ${\bf x}\in \left(\{0,1\}^2\right)^4$ in the following fashion
$$
g_1(u_{(0,1)},u_{(2,3)},u_{(4,5)},u_{(6,7)})={\bf x}_0^3,\,\,\,\,\,\, u_{(2i,2i+1)},x_i\in \{0,1\}^2,\,\,\,i\in[4]_{-}.
$$

The codeword ${\bf x}_0^3$ is transmitted over $4$ copies of a quaternary input memoryless channel $\tilde{\mathcal{W}}$, and  the output vector $\bf y$ is received. By the channel splitting principle we derive the following channels
\begin{description}
  \item[channel $\tilde{\mathcal{W}}_4^{(2i,2i+1)}$:] $
\tilde{W}_4^{(2i,2i+1)}({\bf y},{\bf u}_{0}^{2i-1}|u_{(2i,2i+1)})\triangleq
\sum_{{\bf u}_{2i+2}^7\in \{0,1\}^{6-2i}}\frac{1}{4^{3}}\tilde{W}_4({\bf y}|{\bf u}_{0}^{2i-1},u_{(2i,2i+1)},{\bf u}_{2i+2}^7),$
$$
 u_{(2i,2i+1)}\in \{0,1\}^2,\,\,\,\,\, i\in[4]_{-}.
$$
\end{description}
\normalsize

We denote  $g^{(1)}(\cdot)\triangleq g_0(\cdot)$. Constructing a mapping function of  dimension $16$ (denoted by $g^{(2)}(\cdot)$) is done as follows. Let ${\bf u}$ be a binary vector of length $16$.
Define three vectors ${\bf a} \triangleq g_0(u_0,u_{(1,2)},u_3),$
${\bf b} \triangleq g_1(u_{(4,5)},u_{(6,7)},u_{(8,9)},u_{(10,11)})$ and
${\bf c} \triangleq g_0(u_{12},u_{(13,14)},u_{15}).$
Using these definitions we finally have
\begin{equation}\label{eq:g2Constrc}
g^{(2)}({\bf u})=\big[g_0(a_0,b_0,c_0),g_0(a_1,b_1,c_1),
\end{equation}
$$g_0(a_2,b_2,c_2),g_0(a_3,b_3,c_3)\big]. $$
In this construction ${\bf a},{\bf b}$ and ${\bf c}$ are three outer-codes of length of four symbols (the symbols of $\bf a$ and $\bf c$ are bits, and for $\bf b$ these are quaternary symbols). The outer-codes are combined together using the inner mapping $g_0(\cdot)$.

\begin{figure}
\center
  \includegraphics[scale = 0.08]{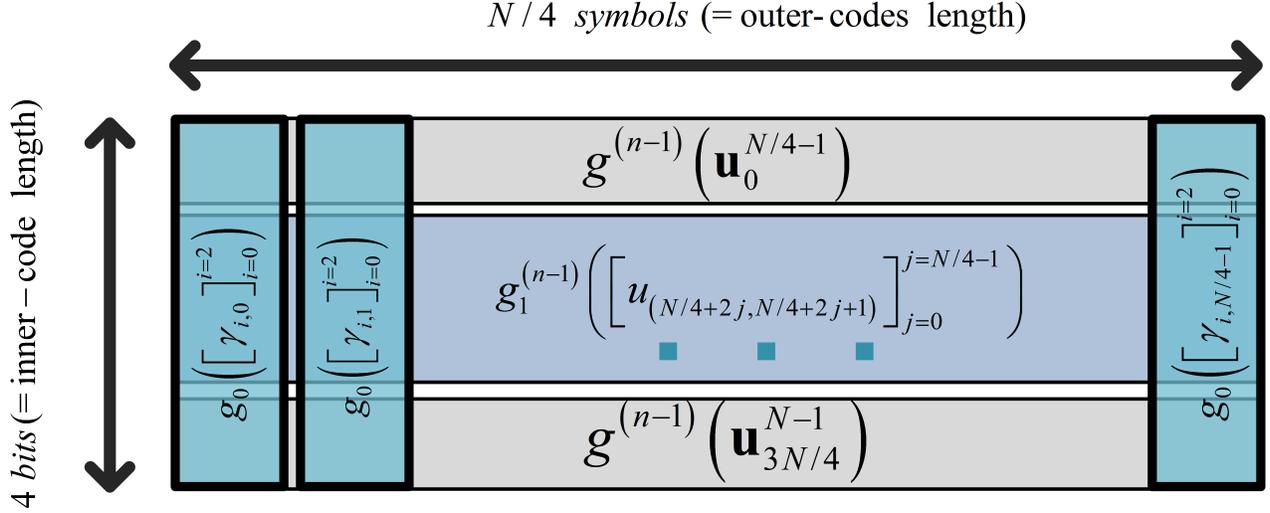}\\
  \caption{A GCC representation of the length $N=4^n$ bits mixed-kernels polar-code $g^{(n)}(\cdot)$  described in Section  \ref{sec:mixKer}   }\label{fig: illustrationOfExampleMixed}
\end{figure}

In order to extend this construction to a mapping $g^{(n)}\left({\bf u}_0^{4^n-1}\right)$, $n>2$ for which some of the inputs are glued together, we suggest the following recursive GCC  construction. We define three outer-code:
\begin{description}
  \item[outer-code  $\#0$:] $\left[\gamma_{0,j}\right]_{j=0}^{j=N/4-1}=g^{(n-1)}\left({\bf u}_0^{N/4-1}\right),\,\,\,\, u_j,\gamma_{0,j}\in \{0,1\} ,\,\,\,\,j\in [N/4]_{-}$.
        \item[outer-code  $\#1$:] $\left[\gamma_{1,j}\right]_{j=0}^{j=N/4-1}=g^{(n-1)}_2\left(\left[{ u}_{\left(N/4+2j,N/4+2j+1\right)} \right]_{j=0}^{j=N/4-1}\right),\,\,\,\, u_{\left(N/4+2j,N/4+2j+1\right)}, \gamma_{1,j} \in \{0,1\}^2 ,\,\,\,\,j\in [N/4]_{-}$.
  \item[outer-code  $\#2$:] $\left[\gamma_{2,j}\right]_{j=0}^{j=N/4-1}=g^{(n-1)}\left({\bf u}_{3N/4}^{N-1}\right),\,\,\,\,u_{3N/4+j},\gamma_{2,j}\in \{0,1\}  ,\,\,\,\,j\in [N/4]_{-}$.
\end{description}
%
Note that outer-codes $\#0$ and $\#2$ are just mixed-kernels constructions of length $N/4$ bits. The encoder of  these outer-codes outputs binary vectors, while its input is described as a mixture of binary and quaternary symbols (generated by bits that were glued together). Outer-code $\#1$ is a  homogenous polar code construction of length $N/4$ quaternary symbols, that  all of its input symbols and output symbols are bits that were glued together in pairs. Finally, these three outer-codes are combined together using the $g_0\left(\cdot\right)$ inner mapping (note the consistency of this definition with that of $g^{(2)}(\cdot)$ in (\ref{eq:g2Constrc})).
\begin{equation}\label{eq:GCCConstructExample}
g^{(n)}=\Big[ g_0\left(\gamma_{0,0}, \gamma_{1,0}, \gamma_{2,0}\right),\,\,g_0\left(\gamma_{0,1}, \gamma_{1,1}, \gamma_{2,1}\right),\ldots,
\end{equation}
$$
\,\,\,\,\,\,\,g_0\left(\gamma_{0,N/4-1}, \gamma_{1, N/4-1}, \gamma_{2, N/4-1}\right)  \Big].
$$
Figure \ref{fig: illustrationOfExampleMixed} depicts the GCC construction of  (\ref{eq:GCCConstructExample}). Note that outer-code $\#1$ was drawn as a rectangle having the same width of  outer-code $\#0$ (or $\#2$). This property symbolizes that all the outer-codes have the same length in terms of \textit{symbols}. On the other hand, the height of the rectangle of outer-code $\# 1$ is twice the height of each of the rectangles of the  other two outer-codes. This property indicates that the symbols alphabet size of  outer-code $\#1$  is twice  the size of the symbols alphabet  of the   other  outer-codes (for which the symbols are bits). This is because outer-code $\#1$ is a quaternary mapping in which both the input symbols and the output symbols are pairs of glued bits.

This mixed-kernels construction can be analyzed using Arikan's \textit{channel tree process} as we later describe in Subsection \ref{sec:treeProcess}. The foundation to this analysis is based on the simple observation that SC decoding of the inputs to the mapping  $g^{(n)}(\cdot)$ is equivalent to decoding inputs to the transformations $g_0(\cdot)$ or $g_1(\cdot)$.  These transformations use as their communication channel one of the  synthetic channels generated by the transformation $g^{(n-1)}(\cdot)$ over the original channel $\mathcal{W}$. In other words, when decoding one bit $u_i$ (two glued bits $u_{(i,i+1)}$) over the channel $W_{4^{n}}^{(i)}({\bf y} ,{\bf u}_0^{i-1}|u_i)$ (over  the channel  $W_{4^{n}}^{(i,i+1)}({\bf y}, {\bf u}_0^{i-1}|u_{(i,i+1)})$), this is manifested as decoding a bit (a glued pair of  bits) which is an input to the transformations $g_0(\cdot)$ or $g_1(\cdot)$. These transformations "see" as the communication channel the appropriate synthetic channel ($\mathcal{W}_{4^{n-1}}^{(j)}$ or $\mathcal{W}_{4^{n-1}}^{(j,j+1)}$, depending on the value of  $i$).
 This description of the synthetic channels evolution enables a recursive analysis of the behavior of the SC decoder.
\subsection{The Channel Tree Process}\label{sec:treeProcess}

We now turn to describe the channel tree process corresponding to our example of mixed-kernels  construction. A random sequence $\left\{W_n\right\}_{n\geq 0}$ is
defined such that $W_{n} \in \left\{\mathcal{W}_{4^n}^{(\tau_n(i))} \right\}_{i=0}^{\nu(n)-1}$, where $\nu(n)$ denotes the number of channels (where the glued bits channels are counted as one channel)  and $\tau_n(i)$ denotes the index of the $i^{th}$ channel  ($\tau_n(i)$ is needed because some of the channels correspond to glued bits and therefore have their indexing as a pair of integer numbers\footnote{ In case $\tau_n(i)=(j_0,j_1)$ we denote, for brevity, the channel indicated by it as $\mathcal{W}_{4^n}^{(j_0,j_1)}$, instead of $\mathcal{W}_{4^n}^{((j_0,j_1))}$ as the notation implies. }). For example, for the $\mathcal{W}_{16}$ channel, constructed using the transformation in (\ref{eq:g2Constrc}), we have the number of channels $\nu(2)=10$, where the values of  $\tau_2(\cdot)$ are $\left[\tau_2(i)\right]_{i=0}^{i=9}=\left[0 ,(1,2),3,(4,5),(6,7),(8,9),(10,11),12,(13,14),15\right]$.
We further denote by $\left\{N_n\right\}_{n\geq 0}$ the number of bits at the input of the channel. We therefore have  $N_n=1$ in case we consider  a single bit input channel and $N_n=2$ in case we deal with a channel of glued bits input. We define the channels random sequence recursively.
\begin{equation}\label{eq:chTreeProcessExample}
W_{n+1}=W_n^{(B_{n})} \,\,\,\text{for } n\geq0\,;\,\,W_0 = \mathcal{W}, N_0 = 1,
\end{equation}
where $B_n\in \left\{0,(1,2),3,(0,1),(2,3),(4,5),(6,7) \right\}$ and indicates the labels of the synthetic channels defined in Subsection \ref{sec:mixKerConstruct} (note that the pairs of numbers in this set correspond to channels having inputs of  two  glued bits).
Moreover, $W_n^{(B_{n})}$ denotes a synthetic channel $\tilde{\tilde{\mathcal{W}}}^{(B_{n})}_4$ where the basic channel $\tilde{\tilde{\mathcal{W}}}$ is taken as the previous element of the channel  tree  process, i.e. $\tilde{\tilde{\mathcal{W}}}=W_n$. The channel realizations  of $W_n$ is the set of all synthetic channels as defined by traversing the inner layers of the GCC construction (and using the appropriate channel splitting formulae).  It is the object of this sequence definition,  that each of the synthetic channels induced by $g^{(n)}(\cdot)$ will have probability $\mathrm{\rho}$ if its input is binary and $2\cdot\mathrm{\rho}$ if its input is quaternary.
This sequence will enable us to utilize the probabilistic-method to prove properties of the polar coding scheme.

The description of the probabilistic dynamics of  the random sequences $\left\{B_n\right\}_{n\geq 0},\left\{N_n\right\}_{n\geq 0}$  now follows. Let $\left\{B_{n}^{(1)}\right\}_{n\geq 0}$ be an i.i.d random sequence of the values $\left[0,(1,2),3\right]$ with corresponding probabilities $\left[0.25,0.5,0.25\right]$, and let $\left\{B_{n}^{(2)}\right\}_{n \geq 0}$ be an i.i.d random sequence of the values $\left[(0,1),(2,3),(4,5),(6,7)\right]$ with uniform probabilities. Denote by the random variable $T$ the minimum non-negative $n$ such that $B_{n}^{(1)}= (1,2)$, and set
$$
N_n =\left\{
       \begin{array}{ll}
         1, & \hbox{$n\leq T$;} \\
         2, & \hbox{$n >  T$.}
       \end{array}
     \right.
$$
Finally, set $B_n=B^{(N_{n})}_n$. Note that $T$ is a geometric random variable with probability of success $p=1/2$. Furthermore, given the value of $T$,  the sequence of $B_n$ is of independent samples (although the distribution is not identical for all samples).

Suppose we have  a certain channel $\mathcal{W}$  and a binary i.i.d input vector $U_0^3$ that is transformed by $g_0(\cdot)$ to $X_0^3$, transmitted over a B-MC channel, and received as $Y_0^3$. The mutual information chain rule implies that
\begin{equation}\label{eq:infAdd}
4\cdot I(\mathcal{W})=I(Y_0^3;U_0^3)=I(Y_0^3;U_0)+I(Y_0^3;U_{1,2}|U_0)+
\end{equation}
$$+I(Y_0^3,U_{3}|U_0^2)=I(\mathcal{W}_4^{(0)})+I(\mathcal{W}_4^{(1,2)})+I(\mathcal{W}_4^{(3)}).$$
\normalsize
Next, define the information random sequence corresponding to the channels as $\left\{I_n\right\}_{n\geq 0 }$.
\begin{equation}\label{eq:InfRV}
I_n = \frac{I(W_n)}{N_n} \,\,\,\,\, n\geq 0.
\end{equation}
For a channel $\mathcal{W}$ with input $X\in \mathcal {X}$ and output $Y\in \mathcal{Y}$, we denote by $P_e\left(\mathcal{W}\right)$ the average error probability of the maximum a posteriori estimator $\hat{x}(y) = \arg\max_{x\in \mathcal X} \Pr\left(X=x|Y=y\right)$. This means that
\begin{equation}
P_e(\mathcal{W}) = 1- \sum_{y\in\mathcal{Y}} \Pr\left(Y=y\right)\cdot \max_{x\in\mathcal{X}}\Pr\left({X=x|Y=y}\right).
\end{equation}
We define the random sequence $P_{e,n}=P_e\left(W_n\right)$.
The \bhat parameter sequence is  denoted by  $Z_n = Z(W_n)$, where
for a $q$-ary channel $\mathcal{W}$ we have $Z(\mathcal{W})=\frac{1}{q\cdot(q-1)}\sum_{x,x'\in \mathcal{X}^2,x\neq x'}Z_{x,x'}\left(\mathcal{W}\right)$  and $$Z_{x,x'}(\mathcal{W})=\sum_{y\in\mathcal{Y}}\sqrt{W\left(y|x\right)W\left(y|x'\right)}.$$


Note that $I_n,Z_n \in [0,1]$. By using \cite[Proposition 3]{Sasog}, it can be shown that $Z_n \rightarrow 1 \iff I_n \rightarrow 0$, and that  $Z_n \rightarrow 0 \iff I_n \rightarrow 1$.
\begin{proposition}\label{propo:infoSeqMartingale}
The process $\left\{I_n\right\}_{n\geq 0 }$ is a bounded martingale which is uniformly integrable. As a result, it converges almost surely to $I_{\infty}$.
\end{proposition}
\proof
Employing the   information sequence definition (\ref{eq:InfRV}) results in

\begin{equation}\label{eq:Mart1}
\mathbb{E}\left[I_{n+1}|I_n,N_n=1\right]=\frac{1}{4}\frac{I\left(W_n^{(0)}\right)}{1}+\frac{2}{4}\frac{I\left(W_n^{(1,2)}\right)}{2}+\frac{1}{4}\frac{I\left(W_n^{(3)}\right)}{1}.
\end{equation}

Using (\ref{eq:infAdd}) we have
$$
\mathbb{E}\left[I_{n+1}|I_n,N_n=1\right]=
$$
\begin{equation}\label{eq:Mart2}
=\frac{1}{4}\left(I\left(W_n^{(0)}\right)+I\left(W_n^{(1,2)}\right)+I\left(W_n^{(3)}\right)\right)=\left.\frac{I(W_n)}{N_n}\right]_{N_n=1}=I_n.
\end{equation}

On the other hand
$$
\mathbb{E}\left[I_{n+1}|I_n,N_n=2\right]=
$$
\begin{equation}\label{eq:Mart3}
=\frac{1}{4}\frac{I\left(W_n^{(0,1)}\right)}{2}+\frac{1}{4}\frac{I\left(W_n^{(2,3)}\right)}{2}+\frac{1}{4}\frac{I\left(W_n^{(4,5)}\right)}{2}+\frac{1}{4}\frac{I\left(W_n^{(6,7)}\right)}{2}
\end{equation}
\normalsize
which is
\begin{equation}\label{eq:Mart4}
\mathbb{E}\left[I_{n+1}|I_n,N_n=2\right]=\frac{1}{2}\cdot\frac{1}{4}\left(I\left(W_n^{(0,1)}\right)+I\left(W_n^{(2,3)}\right)+\right.
\end{equation}
$$
\left.+I\left(W_n^{(4,5)}\right)+I\left(W_n^{(6,7)}\right)\right)=\left.\frac{I(W_n)}{N_n}\right]_{N_n=2}=I_n
$$
\normalsize
Consequently, by taking (\ref{eq:Mart2}) and (\ref{eq:Mart4}) we have
\begin{equation}\label{eq:Mart5}
\mathbb{E}\left[I_{n+1}|I_n\right]=I_n,
\end{equation}
which means that the sequence $\left\{I_n\right\}_{n\geq 0}$ is a martingale. Furthermore, it is uniformly integrable (see e.g. \cite[Theorem 4.5.3]{Chung2001})  and therefore it converges almost surely to $I_{\infty}$.
%
\qed

Note that for any $S \subseteq \mathbb{R}$
\begin{equation}\label{eq:probMethIn}
    \Pr\left(I_n \in S \right)=\frac{1}{4^n}\sum_{i\in [\nu(n)]_{-} \,\,\, s.t. \,\,\, I\left(\mathcal{W}_{4^n}^{\left(\tau_n(i)\right)}\right)\in S}\#\left(\tau_n(i)\right),
\end{equation}
where $\#\left(\tau_n(i)\right)$ counts the number of bits at the input of channel $\tau_n(i)$, which is $1$ for a single bit input channel, and $2$ for a glued two bits input channel.
Observe that (\ref{eq:probMethIn}) attributes to the two bits of the glued bits pair the same characterizations  of  mutual information (because they are regarded as a unified entity), and as such they are counted. Note further that
$
\mathbb{E}\left[I_n\right]=\mathbb{E}\left[I_{\infty}\right]=I(\mathcal{W}).
$ Thus, by showing that the mixed kernels construction is polarizing, i.e. $I_{\infty}\in \{0,1\}$, we may infer using (\ref{eq:probMethIn}) that the proportion of clean channels (induced by the transformation and the
SC decoding) is $I(\mathcal{W})$.

Let  $\Gamma_n$ be the number of glued two bits input channels of  $g^{(n)}(\cdot)$. Using the above probabilistic-method, we can deduce that
\begin{equation} \label{eq:Gamma}
\Gamma_n  = 4^n\cdot \frac{1}{2} \cdot \Pr(N_n = 2) = \frac{4^n}{2}\cdot\left(1- \frac{1}{2^n}\right).
\end{equation}
The proportion of the  glued two bits channel goes to $1$ as $n$ grows, and so is the number of occurences of the $g_1(\cdot)$ kernel. Because of this we refer to  $g_1(\cdot)$ as the \textit{surviving kernel} of the mixed-kernels construction.
As a consequence the properties of $g_1(\cdot)$ dominate the construction asymptotically. Specifically, we show in the sequel,  that if the kernel $g_1(\cdot)$ is polarizing, so is the mixed-kernels construction. Moreover, the polar coding exponent associated with the  $g_1(\cdot)$ kernel also defines the rate of polarization of the mixed-kernels configuration.
\subsection{Polarization and Polarization Rate}\label{sec:PolAndPolRate}
In this part we study the polarization property of our mixed-kernels example and its rate of polarization.
We show that $g_1(\cdot)$'s characteristics determine the attributes of the mixed-kernels structure for asymptotically long codes.
\begin{proposition}\label{propo:plrForMxdKernel}
Assume that $g_1(\cdot)$ is a polarizing kernel, i.e. for a construction that is based only on $g_1(\cdot)$ we
have that
\begin{equation}\label{eq:prPlrForMxKern1}
\lim_{n \rightarrow \infty}\Pr\left(I\left(\tilde{W}_n\right)/2 \in (\delta,1-\delta) \right)=0,\,\,\,\, \forall \delta \in (0,0.5),
 \end{equation}
 where $\left\{\tilde{W}_n\right\}_{n\geq 0}$ is the channel tree process associated with $g_1(\cdot)$.
 As a result, the mixed-kernels construction is also polarizing, i.e.
\begin{equation}\label{eq:prPlrForMxKern2}
\lim_{n \rightarrow \infty}\Pr\left(I_n \in (\delta,1-\delta) \right)=0,\,\,\,\, \forall \delta \in (0,0.5)
 \end{equation}
\end{proposition}

\proof
We prove that for a given $\delta\in(0,0.5)$ for each $\epsilon>0$ there exists an
$n_0=n_0(\delta,\epsilon)$, such that for all $n>n_0$
$$
\Pr\left(I_n \in (\delta,1-\delta) \right)<\epsilon.
$$
Let $n_1$ be chosen such that $\Pr\left(N_n=2\right)\geq 1-\frac{\epsilon}{2}$ for every $n\geq n_1$.
 Now, for $n=n_1$ consider all the channels, $\mathcal{W}_{4^{n_1}}^{(i,j)}$, having glued bits input.
By our assumption, when $n$ grows further, each one of them undergoes polarization. According to (\ref{eq:prPlrForMxKern1}) this means that  each one of the
 $\Gamma_{n_1}$ glued channels has an index $n_2(i,j)$ such that when $n\geq n_1+n_2$
$$
\Pr\left(I(W_{n})/2 \in (\delta,1-\delta) \left| W_{n_1}= \mathcal{W}_{4^{n_1}}^{(i,j)}\right.\right)<\frac{\epsilon}{2}.
$$
Denote by $n_2^{*}$ the maximum over these $n_2(i,j)$, and by $n_0 \triangleq n_1+n_2^*$. We have that for $n\geq n_0$
$$
\Pr\left(I_n \in (\delta,1-\delta) \right) = $$
\begin{equation}\label{eq:fullProb1}
=\underbrace{\Pr\left(I_n \in (\delta,1-\delta) |N_n=1 \right)}_{\leq 1}\underbrace{\Pr\left(N_n=1\right)}_{< \epsilon/2}+\underbrace{\Pr\left(I_n \in (\delta,1-\delta) |N_n=2 \right)}_{<\epsilon/2}\underbrace{Pr\left(N_n=2\right)}_{\leq 1} < \epsilon.
\end{equation}
\normalsize
\qed

We now turn to discuss the polarization rate. In order to do this, we need to consider the partial distances of the kernels.
We use the notations of Mori and Tanaka \cite{MoriandTanka}. For a given kernel $g(v_0,v_1,\ldots,v_{m-1})$ as defined in (\ref{eq:gMixedDefined}), we give the following definitions.
$$
D^{(i)}_{x,x'}\left({\bf v}_{0}^{i-1}\right)=
\min_{{\bf w}_{i+1}^m , {\tilde{\bf w }}_{i+1}^m}d_H\left(g\left({\bf v}_{0}^{i-1},x,{\bf w}_{i+1}^{m-1}\right),g\left({\bf v}_{0}^{i-1},x',\tilde{{\bf w}}_{i+1}^{m-1}\right)\right)
$$
$$
D^{(i)}_{x,x'}=\min_{{\bf v}_{0}^{i-1}}D^{(i)}_{x,x'}\left({\bf v}_{0}^{i-1}\right)\,\,\,x,x'\in F^{\eta_i}
$$
$$
D^{(i)}_{\text{max}}=\max_{x,x'\in F^{\eta_i}}D^{(i)}_{x,x'}\,\,;\,\,D^{(i)}_{\text{min}}=\min_{x,x'\in F^{\eta_i},x\neq x'}D^{(i)}_{x,x'}
$$
\normalsize
In order to distinguish between the partial distances of the two kernels, $g_0(\cdot)$ and $g_1(\cdot)$, we add an additional subscript to these parameters for kernel indication. For example, $D_{0,\text{min}}^{(i)}$ and $D_{1,\text{min}}^{(i)}$ denote the $i^{\text{th}}$ item in the  minimum partial distance sequences of kernel $g_0(\cdot)$ and kernel $g_1(\cdot)$, respectively. We note here that for  linear kernels, we have $D^{(i)}_{\text{max}}=D^{(i)}_{\text{min}}$.

\begin{proposition}\label{propo:exponentEx}
If $g_1(\cdot)$ is a linear polarizing kernel and $Z(\mathcal{W})\neq 0$  then it holds for any $\delta>0$
\begin{equation}\label{eq:exponentEx1_Mod}
\lim_{n\rightarrow \infty}\Pr\left( P_{e,n} \leq 2^{-4^{n(E_c(g_1)-\delta)}}\right)\geq I(\mathcal{W}),
\end{equation}
\begin{equation}\label{eq:exponentEx2_Mod}
\lim_{n\rightarrow \infty}\Pr\left( P_{e,n} \leq 2^{-4^{n(E_c(g_1)+\delta)}}\right)=0.
\end{equation}
where $ {E_c(g_1)}=1/4\sum_{i=0}^3\log_4\left(D_{1,\text{min}}^{(i)}\right)$ and referred to as the exponent of the kernel $g_1(\cdot)$.
\end{proposition}
\proof
Let $\epsilon>0$. Similarly to Proposition \ref{propo:plrForMxdKernel}, we let $n_1$ be chosen such that $\Pr\left(N_n=2\right)\geq 1-\frac{\epsilon}{2}$ for each $n\geq n_1$.
 Now, for $n=n_1$ consider all channels with  glued bits inputs,  $\mathcal{W}_{4^{n_1}}^{(i,j)}$.
 According to Mori and Tanaka \cite[Theorem 31]{Mori2014}, when $n$ grows further, each one of them undergoes polarization and have its error probability  decaying according to the exponent of $g_1(\cdot)$. This means that  each one of the $\gamma_{n_1}$ glued channels has an index $n_2=n_2(i,j,\delta,\epsilon)$ such that for $n\geq n_1+n_2$
\begin{equation}\label{eq:rateOfPlr1}
\Pr\left(  P_{e,n}< 2^{-4^{(n-n_1)\left(E_c(g_1)-\delta/2\right)}} \left| W_{n_1}= \mathcal{W}_{4^{n_1}}^{(i,j)}\right.\right)\geq I\left( \mathcal{W}_{4^{n_1}}^{(i,j)}\right)/2-\epsilon/2,
\end{equation}
\begin{equation}\label{eq:rateOfPlr2}
\Pr\left( P_{e,n}< 2^{-4^{(n-n_1)(E_c(g_1)+\delta)}} \left| W_{n_1}= \mathcal{W}_{4^{n_1}}^{(i,j)}\right.\right)\leq \epsilon/2.
\end{equation}
Here $I\left( \mathcal{W}_{4^{n_1}}^{(i,j)}\right)$ is divided by the number of bits at the input of the channel (which is $2$ because we consider glued bits channels) in accordance to \cite{Mori2014}.
For given $\delta$ and $\epsilon >0$, denote by $n_2^{*}$ the maximum over the aforementioned $n_2(i,j,\delta,\epsilon)$. Also denote by $n_0$ the minimum natural number that is $\geq n_1+n_2^*$ and also satisfies (\ref{eq:req1Forn2}) for all $n\geq n_0$.
\begin{equation}\label{eq:req1Forn2}
    n\left(1-n_1/n\right)\cdot (E_c(g_1)-\delta/2)\geq n\cdot(E_c(g_1)-\delta).
\end{equation}
Therefore, for $n\geq n_0$ we have
\begin{equation}\label{eq:exponentLB_1}
\Pr\left(P_{e,n}< 2^{-4^{n(E_c(g_1)-\delta)}} \left| W_{n_1}= \mathcal{W}_{4^{n_1}}^{(i,j)}\right.\right)\geq I\left( \mathcal{W}_{4^{n_1}}^{(i,j)}\right)/2-\epsilon/2
\end{equation}
and
\begin{equation}\label{eq:exponentUB_1}
\Pr\left(P_{e,n}< 2^{-4^{n(E_c(g_1)+\delta)}} \left| W_{n_1}= \mathcal{W}_{4^{n_1}}^{(i,j)}\right.\right)\leq \epsilon/2.
\end{equation}
Using the law of total probability,

$$
\Pr\left(P_{e,n}< 2^{-4^{n(E_c(g_1)-\delta)}} \right) \geq
$$
$$
\geq \sum_{ i \in \left[\nu(n_1)\right]_{-} \bigwedge |\tau_{n_1}(i)|=2}\Pr\left(P_{e,n}< 2^{-4^{n(E_c(g_1)-\delta)}} \left| W_{n_1}= \mathcal{W}_{4^{n_1}}^{\tau_{n_1}(i)}\right.\right)\cdot \Pr\left( W_{n_1}= \mathcal{W}_{{4^{n_1}}}^{\tau_{n_1}(i)}\right).
$$
Note that here we bound the summation from below by enumerating only the channels which are derived from the glued bits channels in layer $n_1$. Theses channels are identifyied by pair of indices (i.e. $|\tau_{n_1}(i)|=2$). Using (\ref{eq:exponentLB_1}), we derive that
$$
\Pr\left(P_{e,n}< 2^{-4^{n(E_c(g_1)-\delta)}} \right) \geq  \sum_{ i \in \left[\nu(n_1)\right]_{-}  \bigwedge|\tau_{n_1}(i)|=2} \Pr\left( W_{n_1}= \mathcal{W}_{{4^{n_1}}}^{\tau_{n_1}(i)}\right)\cdot \left( I\left( \mathcal{W}_{{4^{n_1}}}^{\tau_{n_1}(i)}\right)/2-\epsilon/2\right)=
$$
$$
=  \sum_{ i \in \left[\nu(n_1)\right]_{-}  \bigwedge |\tau_{n_1}(i)|=2} \Pr\left( W_{n_1}= \mathcal{W}_{4^{n_1}}^{\tau_{n_1}(i)}\right)\cdot I\left( \mathcal{W}_{4^{n_1}}^{\tau_{n_1}(i)}\right)/2-\Pr(N_{n_1}=2)\cdot\epsilon/2=
$$
\small
$$
=  \sum_{  i \in \left[\nu(n_1)\right]_{-}  } \Pr\left( W_{n_1}= \mathcal{W}_{4^{n_1}}^{\tau_{n_1}(i)}\right)\cdot \frac{ I\left( \mathcal{W}_{4^{n_1}}^{\tau_{n_1}(i)}\right)}{|\tau_{n_1}(i)|}\,\,-\sum_{ i \in \left[\nu(n_1)\right]_{-} \bigwedge|\tau_{n_1}(i)|=1} \Pr\left( W_{n_1}= \mathcal{W}_{4^{n_1}}^{\tau_{n_1}(i)}\right)\cdot I\left( \mathcal{W}_{4^{n_1}}^{\tau_{n_1}(i)}\right)-\Pr(N_{n_1}=2)\cdot\epsilon/2\geq
$$
\normalsize
$$
\geq  \mathbb{E}(I_{n_1}) - (1-\Pr(N_{n_1}=2)) -\Pr(N_{n_1}=2)\cdot\epsilon/2\geq I(\mathcal{W}) -\epsilon
$$
On the other hand,
$$
\Pr\left(P_{e,n}< 2^{-4^{n(E_c(g_1)+\delta)}}\right)\leq
$$
$$
\leq \sum_{ i \in \left[\nu(n_1)\right]_{-} \bigwedge|\tau_{n_1}(i)|=2}\Pr\left(P_{e,n}< 2^{-4^{n(E_c(g_1)+\delta)}} \left| W_{n_1}= \mathcal{W}_{4^{n_1}}^{\tau_{n_1}(i)}\right.\right)\cdot \Pr\left( W_{n_1}= \mathcal{W}_{4^{n_1}}^{\tau_{n_1}(i)}\right)+
$$
$$
+\sum_{ i \in \left[\nu(n_1)\right]_{-}  \bigwedge|\tau_{n_1}(i)|=1}\Pr\left( W_{n_1}= \mathcal{W}_{4^{n_1}}^{\tau_{n_1}(i)}\right)\leq
$$
$$
\leq \epsilon/2\cdot\Pr(N_{n_1}=2)+ \Pr(N_{n_1}=1)\leq \epsilon.
$$
\qed

Note that Proposition \ref{propo:exponentEx} is encouraging, because typically the exponent of the auxiliary kernel can be larger than the exponent of the initial kernel. For example, the exponent of  $g_0(\cdot)$ is $0.5$, while the exponent of $G_{RS}(4)$ is $0.573120$.
\section{General Mixed-Kernels}\label{sec:GenAndConc}
Section \ref{sec:mixKer} introduced a specific instance of the mixed-kernels family, formed by two constituent kernels of $\ell=4$ dimensions  and alphabet sizes of $2$ and $4$. In this section we broaden the ideas and techniques of Section \ref{sec:mixKer} to general mixed-kernels schemes.

Consider a case of a mixed-kernels code  over alphabet $F$ of length $N=\ell^n$ $F$-symbols. Let us assume that we have a code decomposition  of the  $F^{\ell}$ space. An $\ell$ dimensions kernel $g_0(\cdot)$ over $F$ can be associated to this decomposition (see Subsection \ref{subsect:kernelAndCodeDecompose}) acting as an inner-code of our GCC construction. Decomposition steps that induce partitioning of the remaining space (defined by the preceding decomposition steps)  to $|F|$ sub-codes are represented by $F$-symbol inputs to $g_0(\cdot)$. Outer-codes of the same mixed-kernels scheme of length $N/\ell$ $F$-symbols are associated with these inputs. On the other hand, decomposition steps that induce partitioning  of  the remaining space to $|F|^{\eta}$ sub-codes ($\eta>1$) are represented by $\eta$  $F$-symbols that are glued together. The interpretation  of gluing  $\eta$  $F$-symbols is that these symbols are decoded as a unified entity by the SC algorithm. In order to meet this decoding specification we employ a length $N/\ell$ outer-code over alphabet $F^{\eta}$. This outer-code's  $F^{\eta}$-symbols are connected to  the inputs of $g_0(\cdot)$ instances  that are associated with this decomposition step. Typically this outer-code is taken to be a homogenous polar code of length $N/\ell$  $F^{\eta}$-symbols constructed by an $\ell$ dimensions kernel over $F^{\eta}$.

We now turn to formalize this generalization. Let $g_0(v_0,v_1,...,v_{m-1})$ be equal to $g(\cdot)$ in (\ref{eq:gMixedDefined}). Denote the set of indices corresponding to glued symbols at the input of $g_0(\cdot)$ by $\mathcal{B}=\left\{i\in[m]_{-}|\eta_i\geq 2\right\}$ and let $\theta_i\triangleq\sum_{k=0}^i \eta_k$ for $i\in [m]_{-}$ and $\theta_{-1}\triangleq 0$. For each $i\in\mathcal{B}$ we assign a kernel $g_{i+1}(\cdot):\left(F^{\eta_i}\right)^{\ell}\rightarrow \left(F^{\eta_i}\right)^{\ell}$ (if $\eta_i=\eta_j$ we usually employ the same kernel, i.e. $g_{i+1}(\cdot)\equiv g_{j+1}(\cdot)$). The kernels mentioned here are called the \textit{constituent kernels} of the construction. The mapping $g_0(\cdot)$ is referred to as the \textit{interface kernel} and the other kernels are dubbed \textit{auxiliary kernels}.
We note that in \cite[Table 5]{LitsynTblBinaryCodes}, the author gives a list of code  decompositions
that can be used for the definition of a binary interface kernel $g_0(\cdot)$.
Mori and Tanaka's non-binary kernels \cite{MoriandTanka3} may be found suitable for the auxiliary kernels $g_{i+1}(\cdot)$, $i\in \mathcal{B}$.

The construction of a high dimensions transform of length $N=\ell^n$ $F$-symbols, $g^{(n)}\left({\bf u}_{0}^{\ell^{n-1}}\right)$, can be exercised by a proper adjustment of the recursive GCC method we described in Section \ref{sec:mixKer}. For $n=1$ we have $g^{(1)}(\cdot) \equiv g_0(\cdot)$. For $n>1$  we employ the auxiliary kernels  $g_{i+1}(\cdot)\,\,\,\,i\in\mathcal{B}$ which support  the glued symbols inputs of the inner-mapping, $g_0(\cdot)$. Specifically, for length $N=\ell^n$ $F$-symbols code, we have $m$ outer-codes of  length of $N/{\ell}$ symbols (these symbols may be produced by gluing together several $F$-symbols). We denote outer-code $\#i$ by the vector $\left[ \gamma_{i,j}\right]_{j=0}^{N/\ell -1}$, where $i\in [m]_{-}$. Denote by $s_i$ the input offset for outer-code $\#i$, which means that the first index of the input vector $\bf u$ to outer-code $\#i$ is $s_i$. Consequently, we have $s_i = \theta_{i-1}\cdot N / \ell$. If $\eta_i=1$ then the outer-code is an instance of the same mixed-kernels structure of length $N/{\ell}$  $F$-symbols:
\begin{equation}
\left[\gamma_{i,j}\right]_{j=0}^{j=N/\ell-1}=g^{(n-1)}\left(\left[u_{s_i+j} \right]_{j=0}^{j=N/\ell-1}\right),\,\,\,\, u_{s_i+j},\gamma_{i,j} \in F ,\,\,\,\,j\in \left[N/\ell\right]_{-}.
\end{equation}
If $\eta_i > 1$ it means that $v_i$ is a glued symbol of $\eta_i$ $F$-symbols, with corresponding kernel $g_{i+1}(\cdot)$. Therefore, outer-code $\#i$ is an instance of a polar code of length $N/{\ell}$ of $F^{\eta_i}$-symbols. This code is generated by using the homogenous kernel $g_{i+1}(\cdot)$. Formally, we have
\begin{equation}\label{eq:genMixedDef2}
\left[\gamma_{i,j}\right]_{j=0}^{j=N/\ell-1}=g^{(n-1)}_{i+1}\left(\left[u_{\left(s_i+\eta_i\cdot j\,,\,s_i+\eta_i\cdot (j+1)-1\right)} \right]_{j=0}^{j=N/\ell-1}\right),\,\,\,\,{ u}_{\left(s_i+\eta_i\cdot j\,,\,s_i+\eta_i\cdot (j+1)-1\right)}, \gamma_{i,j} \in F^{\eta_i} ,\,\,\,\,j\in [N/\ell]_{-}.
\end{equation}
Note that in (\ref{eq:genMixedDef2}), the argument of $g_{i+1}^{(n-1)}(\cdot)$ is a vector of length $N/\ell$. Each element of this vector, $u_{\left(s_i+\eta_i\cdot j\,,\,s_i+\eta_i\cdot (j+1)-1\right)}$, is constructed by $\eta_i$  $F$-symbols,  ${\bf u}_{s_i+\eta_i\cdot j}^{s_i+\eta_i\cdot (j+1)-1}$, that are glued together.
 Finally, these $m$ outer-codes are combined together using the $g_0$ inner mapping
$$g^{(n)}=\Big[ g_0\left(\gamma_{0,0}, \gamma_{1,0},\ldots, \gamma_{m-1,0}\right),\,\,g_0\left(\gamma_{0,1}, \gamma_{1,1},\ldots,  \gamma_{m-1,1}\right),\ldots,$$
$$
\,\,\,\,\,\,\,g_0\left(\gamma_{0,N/\ell-1}, \gamma_{1, N/\ell-1},\ldots,  \gamma_{m-1, N/\ell-1}\right)  \Big].
$$

 Assume that ${\bf x}_0^{\ell-1}=g_0(v_0,v_1,...,v_{m-1})$ is transmitted over $\ell$ copies of the memoryless channel $\mathcal{W}$, and we receive the output vector $\bf y$. The channel splitting principle dictates the generation of $m$ synthetic channels. If the input of channel $\# i$ is over $F$ (i.e. not glued), then we denote the channel by $\mathcal{W}_\ell^{(\theta_{i-1})}$ and we have the following transition function.
\begin{equation}\label{eq:genMxdKernelNonGluedChannel}
{W}_\ell^{(\theta_{i-1})}({\bf y},{\bf u}_0^{\theta_{i-1}-1}|u_{\theta_{i-1}})=\frac{1}{|F|^{\ell-1}}\cdot\sum_{{\bf u}_{\theta_i}^{\ell-1}\in F^{\ell-\theta_i}}{W}_\ell({\bf y}|{\bf u}_0^{\theta_{i-1}-1},u_{\theta_{i-1}},{\bf u}_{\theta_i}^{\ell-1}),\,\,\,\,\, u_{\theta_{i-1}}\in F.
\end{equation}
Glued symbols are handled as a unified entity in SC decoding. If  the input to channel $\# i$ is of $\eta_i$ glued $F$-symbols then we denote the channel by $\mathcal{W}_\ell^{(\theta_{i-1},\theta_i-1)}$. Note that the superscript that identifies the channel ($\left(\theta_{i-1},\theta_i -1 \right)$ in our case) is a pair of numbers  indicating the range of indices of $\bf u$ that were glued together. We have
\begin{equation}\label{eq:genMixedKernelGluedChannelEq}
{W}_\ell^{(\theta_{i-1},\theta_i-1)}\left({\bf y},{\bf u}_0^{\theta_{i-1}-1}\left|u_{\left(\theta_{i-1},\theta_i-1\right)}\right.\right)=\frac{1}{|F|^{\ell-\eta_i}}\cdot\sum_{{\bf u}_{\theta_i}^{\ell-1}\in F^{\ell-\theta_i}}{W}_\ell\left({\bf y}\left|{\bf u}_0^{\theta_{i-1}-1},u_{\left(\theta_{i-1},\theta_i-1\right)},{\bf u}_{\theta_i}^{\ell-1}\right.\right),\,\,\,\,\, u_{\left(\theta_{i-1},\theta_i-1\right)}\in F^{\eta_i}.
\end{equation}
The processing of the likelihoods related to the kernels $g_{i+1}(\cdot)$ for the glued symbols $i\in \mathcal{B}$ is done over a channel $\tilde{ \mathcal{W}}$ with input symbol $F^{\eta_i}$. $\tilde{ \mathcal{W}}$ can be created as a result of one of the channel splittings that were induced by a glued input $v_i$ of $g_0(\cdot)$ (these channels are denoted by a pair of numbers in their superscript, i.e. $\mathcal{W}_\ell^{(\theta_{i-1},\theta_i-1)}$). $\tilde{ \mathcal{W}}$ can also be produced by the homogenous polar code that is connected to a glued input $v_i$. We denote the synthetic channels that are splitted from $\tilde{\mathcal{W}}$ by $\left\{\tilde{\mathcal{W}}_{\ell}^{\left(j \cdot  \eta_i\,,\, (j+1) \cdot\eta_i-1\right)} \right\}_{j=0}^{\ell-1}$. Formally we have
$$
g_{i+1}\left(u_{\left(0,\eta_i-1\right)},u_{\left(\eta_i,2\eta_i-1\right)},\ldots,u_{\left((\ell-1)\cdot\eta_i,\ell\cdot\eta_i-1\right)}\right)={\bf x}_0^{\ell-1},\,\,\,\,\,\, u_{\left(j \cdot \eta_i \,,\, (j+1)\cdot\eta_i-1\right)},x_j\in F^{\eta_i},\,\,\,j\in[\ell]_{-}.
$$
${\bf x}_0^{\ell-1}$ is transmitted over $\ell$ copies of an $F^{\eta_i}$ input memoryless channel $\tilde{\mathcal{W}}$, and  the output vector $\bf y$ is received. By the channel splitting principle we derive the following synthetic  channels $\text{for } j\in[\ell]_{-}$.
$$
\tilde{W}_{\ell}^{\left( j \cdot\eta_i\,,\, (j+1)\cdot\eta_i-1\right)}\left({\bf y},{\bf u}_{0}^{j\cdot\eta_i-1}\left|u_{\left(j\cdot\eta_i\,,\,(j+1)\cdot\eta_i-1\right)}\right.\right)=
$$
$$=\frac{1}{|F|^{\eta_i(\ell-1)}}\cdot\sum_{{\bf u}_{(j+1)\eta_i}^{\ell\cdot\eta_i -1}\in \left(F^{\eta_i}\right)^{\ell-1-j}}\tilde{W}_{\ell}^{\left(j\cdot\eta_i \,,\, (j+1)\cdot\eta_i-1\right)}\left({\bf y}\left|{\bf u}_{0}^{j\cdot\eta_i-1},u_{\left(j\cdot\eta_i\,,\,(j+1)\cdot\eta_i-1\right)},{\bf u}_{(j+1)\cdot\eta_i}^{\ell\cdot\eta_i -1}\right.\right).$$

\begin{example}\label{ex:l8mixedExample}
Let $\ell=8$ and define the following binary output kernel $g_0\left(\cdot\right)$
$$
    g_0\left(u_0,u_{(1,3)},u_{(4,6)},u_7\right) = {\bf u}_0^7 \cdot G,\,\,\,\, u_i\in \{0,1\},i\in [8]_{-},
$$
where $G$ is $8\times 8$ matrix derived  by swapping row $3$ with row $4$ of  $\left(
                                                                                 \begin{array}{cc}
                                                                                   1 & 0 \\
                                                                                   1 & 1 \\
                                                                                 \end{array}
    \right)^{\bigotimes 3}$, where $A^{\bigotimes k}$ denotes the $k^{\text{th}}$ Kronecker power of the matrix $A$. $g_0(\cdot)$ induces a code decomposition of $\{0,1\}^8$  having the following chain of parameters     $(8,8,1)-(8,7,2)-(8,4,4)-(8,1,8)$.  $g_0(\cdot)$ induce a code decomposition of $\{0,1\}^8$  having the following chain of parameters     $(8,8,1)-(8,7,2)-(8,4,4)-(8,1,8)$.
We therefore have two glued octonary input symbols $u_{(1,3)},u_{(4,6)}\in \{0,1\}^3$, that require additional octonary kernels of $\ell=8$ dimensions. We denote these kernels by $g_2(\cdot)$ and $g_3(\cdot)$, respectively. Note that $g_2(\cdot)$ and  $g_3(\cdot)$ are mappings in $\left(\{0,1\}^3\right)^8\rightarrow \left(\{0,1\}^3\right)^8$. We may choose to use the $G_{RS}\left(8\right)$ kernel \cite{Mori2010} both for $g_2(\cdot)$ and for $g_3(\cdot)$. Figure \ref{fig: GCCMixed8} illustrates the GCC construction of length $N=8^n$ bits polar code using this mixed structure.
\begin{figure}
\center
  \includegraphics[scale = 0.15]{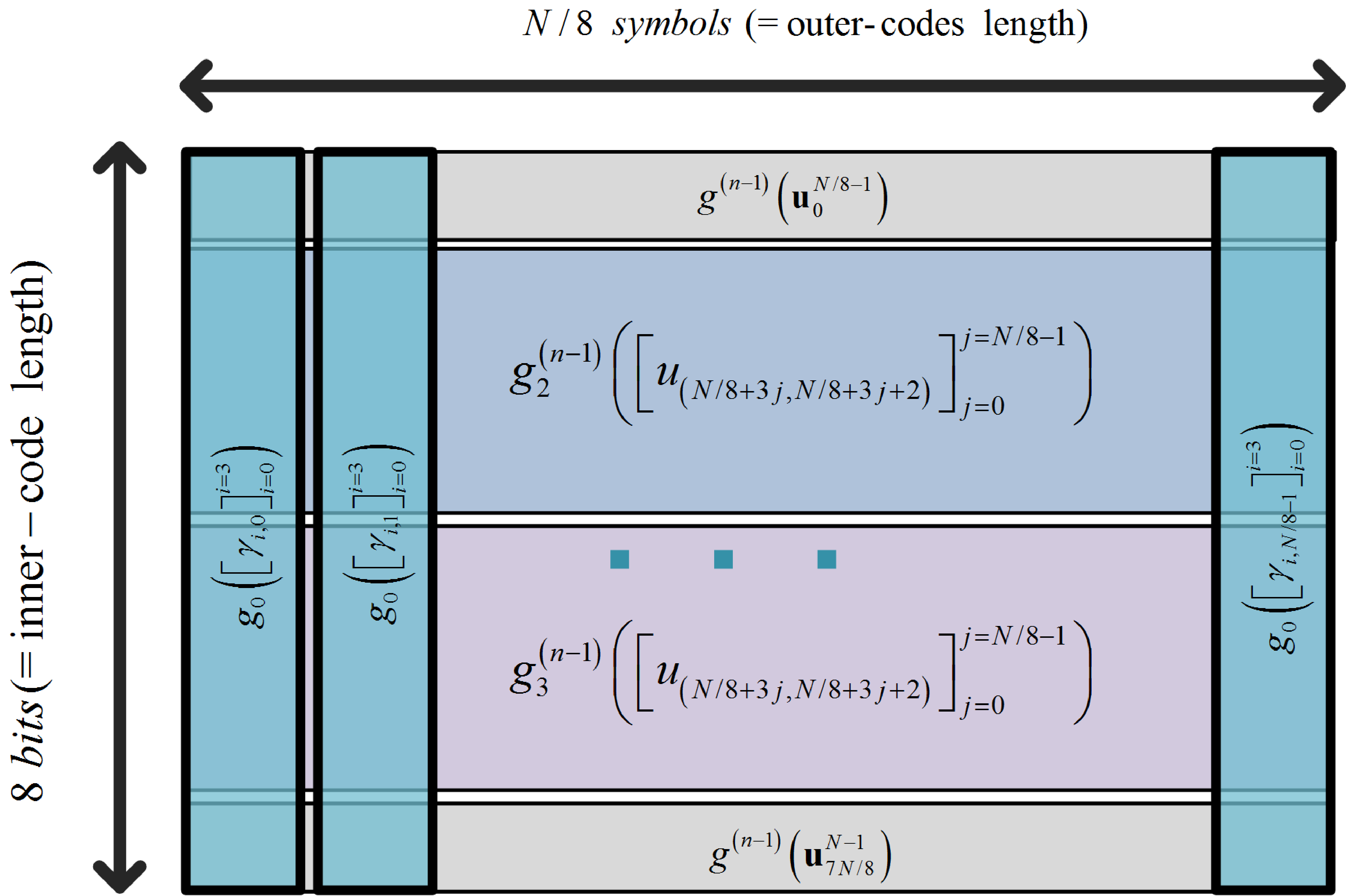}\\
  \caption{A GCC representation of Example  \ref{ex:l8mixedExample} mixed-kernels structure, for length $N=8^n$ bits code (defined by the $g^{(n)}(\cdot)$ mapping). The mapping $g^{(n-1)}(\cdot)$ is the same mixed-kernels construction of length $N/8$ bits, $g_2^{(n-1)}(\cdot)$ and $g_3^{(n-1)}(\cdot)$ are mappings of homogenous polar codes of length $N/8$ octonary symbols.     }\label{fig: GCCMixed8}
\end{figure}
\end{example}

We associate to the mixed-kernels construction a channel tree process,  $W_{n} \in \left\{\mathcal{W}_{\ell^n}^{(\tau_n(i))} \right\}_{i=0}^{\nu(n)-1}$, where $\nu(n)$ denotes the number of synthetic channels induced by the length $\ell^n$ mapping (where  glued symbols input channels are counted as one channel). Moreover, similarly to the definitions in Subsection \ref{sec:treeProcess}, $\tau_n(i)$ denotes the index of channel number $i$. As before, $\left\{N_n\right\}_{n\geq 0}$ denotes the number of symbols at the input of the channel, which in our case is $N_n=1$ when we deal with a single symbol channel or $N_n=\eta_i,\,\,\,\, i\in \mathcal{B}$ when we consider  a channel with input of glued symbols. The channel tree process statistics is defined as follows.
$$
W_{n+1}=W_n^{(B_{n})} \,\,\,\text{for } n\geq0\,;\,\,W_0 = \mathcal{W}, N_0 = 1,
$$
where
\begin{equation}\label{eq:BranchingProcessGen}
B_n =\left\{
       \begin{array}{ll}
          B_n^{(0)}, & \hbox{$n\leq T$;} \\
         B_n^{(i+1)}, & \hbox{$n >  T \bigwedge B_{T}^{(0)} =\left(\theta_{i-1},\theta_i-1\right), i\in \mathcal{B}$ }
       \end{array}.
     \right.
\end{equation}
Note that $W_n^{(B_{n})}$ denotes a synthetic channel $\tilde{\tilde{\mathcal{W}}}^{(B_{n})}_{\ell}$ where the basic channel $\tilde{\tilde{\mathcal{W}}}$ is taken as the previous element of the channel  tree  process, i.e. $\tilde{\tilde{\mathcal{W}}}=W_n$. The sequence $B_n$ indicates the branching of the tree process. Pairs of numbers in the sequence of  $B_n$ indicate channels having input of glued symbols, while single numbers correspond to channels with with $F$-symbol input. The sequence begins by taking the values of the sequence  $B_n^{(0)}$ that correspond to the channels generated by the interface kernel $g_0(\cdot)$. Starting from $n>T$, $B_n$ takes the values of the sequence $B_n^{(i+1)}$ corresponding to the chosen auxiliary kernel.

The random sequence $\left\{B_{n}^{(0)}\right\}_{n\geq 0}$ is i.i.d and takes values from the set $\left\{ \theta_{i-1} | i\notin \mathcal{B} \right\}\bigcup\left\{\left(\theta_{i-1},\theta_i-1\right)|i\in \mathcal{B}\right\}$. The left set in the union is the set of channel indices with non-glued input symbols. Each one of these indices has probability of $1/\ell$. The right set in the union is the set of the indices of channels with glued symbols inputs, such that $\left(\theta_{i-1},\theta_i-1\right)$ has probability of $\eta_i/\ell$. Moreover, for each $i\in\mathcal{B}$, let us define $\left\{B_{n}^{(i+1)}\right\}_{n \geq 0}$ to be an i.i.d random sequence of the values $\left\{\left(j\cdot \eta_i,(j+1)\eta_i-1\right)|j\in[\ell]_{-}\right\}$ with uniform probabilities ($=1/\ell$) associated to each one of them.

Denote by the random variable $T$ the minimum non-negative $n$ such that $B_{n}^{(0)}\in \left\{\left(\theta_{i-1},\theta_i-1\right)|i\in \mathcal{B}\right\}$ (i.e. it refers to an index of a synthetic channel induced by $g_0(\cdot)$ with glued symbols input). It is easy to see that the random variable $T$ is geometric with parameter $p=\left(\sum_{i\in \mathcal{B} }\eta_i\right)/\ell$. Furthermore, given the value of $T$ the sequence  $B_n$ is of independent samples. Since we begin our tree process with channels corresponding to the interface kernel $g_0(\cdot)$ the random variable $T$ indicates the index of transition from channels associated with $g_0(\cdot)$ to channels corresponding to the auxiliary kernels. The specific kernel, to which we transition, is determined by the chosen index  in the transition point. Moreover let $N_n$ indicate the number of symbols at the input of the channel $W_n$. If the transition was to kernel of $\eta_i$ glued symbols we have $N_n= \eta_i$ for $n>T$. Formally,
$$
N_n =\left\{
       \begin{array}{ll}
         1, & \hbox{$n\leq T$;} \\
         \eta_i, & \hbox{$n >  T \bigwedge B_{T}^{(0)} =\left(\theta_{i-1},\theta_i-1\right), i\in \mathcal{B}$. }
       \end{array}
     \right.
$$

Let $\phi(\eta_i)$ denote the number of inputs to $g^{(0)}(\cdot)$, having $\eta_i$ glued symbols. For instance, in Example \ref{ex:l8mixedExample} we have $\phi(1) = 2$ and $\phi(3) = 2$. Denote by $\Gamma_n\left(\eta_i\right)$ the number of inputs of $g^{(n)}(\cdot)$ having $\eta_i$ glued symbols.
 We have that for $\eta_i>1$
\begin{equation}
\Gamma_n\left(\eta_i\right) = \ell^n\cdot\frac{1}{\eta_i}\cdot\Pr\left(N_n=\eta_i\right) = \ell^n\cdot\frac{1}{\eta_i}\cdot\sum_{\tau=0}^{n-1}\Pr\left(N_n=\eta_i \bigwedge T=\tau\right) =
\end{equation}
$$
    =  \ell^n\cdot\frac{1}{\eta_i} \cdot\sum_{\tau=0}^{n-1} \frac{\phi(\eta_i)\cdot\eta_i}{\ell} \cdot (1-p)^{\tau} = \ell^{n-1}\cdot\phi(\eta_i)\cdot \frac{1-(1-p)^n}{p}
$$
For $\eta_i  = 1$ we have
\begin{equation}
\Gamma_n\left(1\right) = \ell^n\cdot \Pr\left(T>n\right)=\ell^n\cdot (1-p)^n.
\end{equation}
The number of $g^{(n)}(\cdot)$'s $F$-symbols  inputs  that are part of an $\eta_i$-glued set is $\eta_i\cdot\Gamma_n(\eta_i)$. Observe that asymptotically in $n$, the proportion of $F$-symbols input   that are part of any $\eta_i$-glued input set is $\frac{\phi(\eta_i)\cdot \eta_i}{p\cdot\ell}$. On the other hand, the proportion of input symbols that are not part of any glued symbols set vanishes as $n$ grows, and thus is also the relative number of occurrences of the initial interface kernel in the construction.  Consequently, the auxiliary kernels are also called the \textit{surviving} kernels of the construction.

Let us define the mutual information sequence   as $I_n=\frac{I(W_n)}{N_n}$ (note that we take $|F|$ as the base of the logarithm in the mutual information definition).
As we  demonstrated in Section \ref{sec:mixKer}, here also the polarization and the rate of polarization properties are determined by the surviving kernels. The latter observation on the dominance of the auxiliary kernels will be evident in the generalization of the propositions from  Section \ref{sec:PolAndPolRate} that are presented next.
\begin{proposition}\label{propo:infoSeqMartingaleGen}
The process $\left\{I_n\right\}_{n\geq 0 }$ is a bounded martingale which is uniformly integrable. As a result, it converges almost surely to $I_{\infty}$.
\end{proposition}
\proof
This proof is similar to the proof of Proposition \ref{propo:infoSeqMartingale}. The only delicate step that we need to consider here is the channel splitting due to the kernel $g_0(\cdot)$. This is because the other kernels are homogenous and polarizing and therefore their information sequence being a martingale was already proven (see e.g. \cite[Lemma 9]{Mori2010}). We have

\begin{equation}\label{eq:MartGen1}
\mathbb{E}\left[I_{n+1}|I_n,N_n=1\right]=\sum_{i\notin \mathcal{B}}\frac{1}{\ell}\cdot I\left(W_n^{(\theta_{i-1})}\right)+ \sum_{i\in\mathcal{B}} \frac{\eta_i}{\ell} \cdot \frac{I\left(W_n^{(\theta_{i-1},\theta_i-1)}\right)}{\eta_i} =
\end{equation}
$$
= \frac{1}{\ell}\left(\sum_{i\notin \mathcal{B}}\cdot I\left(W_n^{(\theta_{i-1})}\right)+ \sum_{i\in\mathcal{B}} I\left(W_n^{(\theta_{i-1},\theta_i-1)}\right)\right) = I_n,
$$
where the last transition is due to the mutual-information chain rule.
As a result of the law of total expectation, we have
\begin{equation}\label{eq:Mart5}
\mathbb{E}\left[I_{n+1}|I_n\right]=\mathbb{E}_{N_n}\left[\mathbb{E}\left[I_{n+1}|I_n,N_n\right]\right]=I_n,
\end{equation}
which means that the sequence $\left\{I_n\right\}_{n\geq 0}$ is a martingale. Furthermore, it is uniformly integrable (see e.g. \cite[Theorem 4.5.3]{Chung2001})  and therefore it converges almost surely to $I_{\infty}$.\qed
\begin{proposition}\label{propo:plrForMxdKernelGen}
Assume that for all $i\in \mathcal{B}$, $g_{i+1}(\cdot)$ is a polarizing kernel, i.e. for a construction that is based only on $g_{i+1}(\cdot)$ we
have that $$\lim_{n \rightarrow \infty}\Pr\left(I\left(\tilde{W}_n\right)/\eta_i \in (\delta,1-\delta) \right)=0,\,\,\,\forall\delta \in (0,0.5).$$
Here $\left\{\tilde{W}_n\right\}_{n\geq 0}$ denotes the channel tree process for the homogenous kernel $g_{i+1}(\cdot)$.
As a result, the mixed-kernels construction is also polarizing, i.e.
$$\lim_{n \rightarrow \infty}\Pr\left(I_n \in (\delta,1-\delta) \right)=0,\,\,\,\,\,\forall\delta \in (0,0.5).$$
\end{proposition}
\proof
The proof is similar to the proof of Proposition \ref{propo:plrForMxdKernel}, only that  here $n_1$ is  chosen  such that $\Pr(N_n>1) \geq 1-\epsilon/2$ for each $n\geq n_1$. \qed

Assume that all the construction's auxiliary kernels are linear. Let $E_{min}=\min_{i\in\mathcal{B}}E_c\left(g_{i+1}\right)$ and $E_{max}=\max_{i\in\mathcal{B}}E_c\left(g_{i+1}\right)$, where
$E_c\left(g_{i+1}\right)$ is the polar coding exponent of kernel $g_{i+1}(\cdot)$ (the base of the logarithm in the polar coding exponent is the kernel size, $\ell$).


\begin{proposition}\label{propo:rateForMxdKernelGeneral}
If for all $i\in\mathcal{B}$ we have that $g_{i+1}(\cdot)$ is a linear polarizing kernel and $Z(\mathcal{W})\neq 0$  then it holds for all $\delta>0$
\begin{equation}\label{eq:exponentEx1_Mod}
\lim_{n\rightarrow \infty}\Pr\left( P_{e,n} \leq 2^{-\ell^{n(E_{min}-\delta)}}\right)\geq I(\mathcal{W});
\end{equation}
\begin{equation}\label{eq:exponentEx2_Mod}
\lim_{n\rightarrow \infty}\Pr\left( P_{e,n} \leq 2^{-\ell^{n(E_{max}+\delta)}}\right)=0.
\end{equation}
\end{proposition}
\proof
This proof is similar to the proof of Proposition \ref{propo:exponentEx}, only that here we may have more than one auxiliary kernel. Consequently, as in the previous proof, we choose $n_1$ such that $\Pr(N_n>1)\geq 1-\epsilon/2$. Because $\forall i\in \mathcal{B}, E_{min}\leq E_c(g_{i+1})\leq E_{max}$ and based on Mori and Tanaka \cite[Theorem 31]{Mori2014}, each  one of the glued channels in layer $n_1$ has an index $n_2 = n_2(i,j,\delta,\epsilon)$ such that for $n\geq n_1+n_2$ we have (\ref{eq:rateOfPlrGen1}) and (\ref{eq:rateOfPlrGen2}) replacing (\ref{eq:rateOfPlr1}) and (\ref{eq:rateOfPlr2}), respectively.
\begin{equation}\label{eq:rateOfPlrGen1}
\Pr\left(  P_{e,n}< 2^{-\ell^{(n-n_1)\left(E_{min}-\delta/2\right)}} \left| W_{n_1}= \mathcal{W}_{\ell^{n_1}}^{(i,j)}\right.\right)\geq I\left( \mathcal{W}_{\ell^{n_1}}^{(i,j)}\right)/2-\epsilon/2,
\end{equation}
\begin{equation}\label{eq:rateOfPlrGen2}
\Pr\left( P_{e,n}< 2^{-\ell^{(n-n_1)(E_{max}+\delta)}} \left| W_{n_1}= \mathcal{W}_{\ell^{n_1}}^{(i,j)}\right.\right)\leq \epsilon/2.
\end{equation}
The rest of the proof of Proposition  \ref{propo:exponentEx} may now be employed. \qed

\section{Merits of Mixed-Kernels Constructions}\label{sec:mixedPotAdvantage}
In this section we discuss possible benefits of using  mixed-kernels based structures. Subsection \ref{sec:improvCodeDec} considers the opportunity for utilizing variety of code decompositions, which may be more suitable for SC decoding. Subsection \ref{sec:redDecComplx} examines the SC algorithm complexity, and shows that for mixed-kernels structures we may have smaller SC decoding complexity compared to homogenous polar codes that are based on their auxiliary kernels (although they have the same polar coding exponent). Furthermore, we suggest an approach to  fairly compare  between the error correction performance of different coding schemes of the same length under SC List (SCL) decoder. This idea will be used in Section \ref{sec:simulations} to demonstrate by simulations that  mixed-kernels structures may outperform the  currently known polar coding schemes.
\subsection{Improved Code Decomposition}\label{sec:improvCodeDec}
Our initial motivation for studying  mixed-kernels structures was the opportunity for generating richer classes of code decompositions that induce new polar code types. These decompositions  allow their codes
to be partitioned into variable numbers of subsets on each step. By doing so, we may be able to ensure that each step increases the partial distance between the sub-codes.  The outer-codes that are associated with these steps can be adjusted to support the different qualities of the resultant synthetic channels. As a consequence, the outer-codes performance under SC decoding may be improved.

As an example of this advantage, let us compare the partial distances that are induced by the different steps of several decompositions of length $N=2^n$ bits codes. Table \ref{tbl:partialDistComb} compares the partial distance sequences that were induced by codes based on  Arikan's binary $(u+v,v)$ kernel, homogenous polar code based on $G_{RS}(4)$ kernel and the mixed-kernels construction of Section \ref{sec:mixKer} with $G_{RS}(4)$ as the auxiliary mapping $g_1(\cdot)$ (denoted as $Mixed-RS4$). The  quaternary output symbols of the $RS4$ constructions are interpreted as two-tuples of  bits in the obvious way.

The partial distances appearing at the table were calculated based on Lee and Yang's \cite[Theorem 7]{Lee2014}.
The entries in the distance sequences are sorted in ascending order and do not necessarily correspond to the exact order of the code decomposition steps. Entries with boldface typeface indicate steps of code decomposition into four sub-codes (and as such they are indicated by the quaternary input symbols to the mapping in Definition \ref{def: transAssiciatedTheCodes}).  Light typeface indicates  decompositions into two sub-codes (and as such they are indicated by a binary input to the mapping). For brevity we use the notation $b^{(m)}$ for $b,m$ natural numbers to denote the vector of $m$ copies of the number $b$.

In order to allow comparisons between structures having the same length, we had to introduce  additional inner-code layers to some of the structures. Specifically, for $N\in \left\{8,32\right\}$  we included an additional $(u+v,v)$ layer as an inner mapping to the $Mixed-RS4$ construction. This means that  for a code of length $N$ bits we have two instances  of length $N/2$ bits $Mixed-RS4$  structures  serving as outer-codes. These two codes are combined together using  the $(u+v,v)$ inner-code. We refer to such structures as the $(u+v,v)-Mixed-RS4$ constructions. Similarity, for $N\in\left\{16,64\right\}$ we included an additional quaternary $(u+v,v)$ layer as an inner mapping to the $RS4$ constructions.  We refer to such structures as the $(u+v,v)-RS4$ constructions.

Although the codes considered here are very short, we may use them also to infer on larger length codes. The reasoning behind this statement is that the examples in the table may be considered as inner-codes of larger GCC constructions. Specifically,  for large length $2^n$ bits codes where $n$ is even, the $N\in \left\{16,64 \right\}$ entries in Table \ref{tbl:partialDistComb} may be considered as inner-codes of the GCC construction. Similarly, when $n$ is odd,  the $N\in \left\{8,32 \right\}$ entries are the inner-codes of the constructions.

The table shows  that the partial distance sequences of the $(u+v,v)$ construction contain  more repetitions of values    compared to the $Mixed-RS4$ structure. We therefore may assume that there are many synthetic channels of similar qualities under SC decoding for the $(u+v,v)$ scheme, while the quality of the channels is more diverse for the mixed-kernels construction. This diversity may lead to better adjustment of the outer-codes that "operate" over these synthetic channels.

 The  partial distance sequences of the mixed-kernels structure can also be interpreted as binary partial distance sequence. According to this interpretation, each quaternary decomposition, denoted by bold entry $\bf b$, is transformed into two steps of binary entries $\left(b,b\right)$. When comparing the induced binary distance sequences of the mixed-kernels with the sequence of $(u+v,v)$ we can observe that for $N\in\{16,32\}$, the mixed-kernels structure has better sequence. The meaning of the last statement is that  for each entry $\alpha_i$ and $\beta_i$ of the distance sequences of the $(u+v,v)$ and the $Mixed-RS4$ structures, respectively, we have $\alpha_i\leq\beta_i$ and there exists $i$, such that $\alpha_i<\beta_i$. Moreover, for $N=16$ bits the mixed structure has the same binary partial distance sequence as the one derived by Korda \textit{et al.} \cite[Example 28]{Korada2010}. Korada \textit{et al.} proved that this sequence is optimal  for binary linear kernels of length $\ell=16$ bits. The optimality here is in the sense that it has the maximum polar coding exponent.

The partial distance sequences of the mixed-kernels structure are better than the sequences of the $RS4$ construction (when both of them are interpreted as binary decompositions). On the other hand, the $RS4$ structure has less decomposition steps than the one induced by the other structures because all the steps are quaternary.  Furthermore, the $RS4$  structure's partial distance sequence contains fewer repeating elements than the other schemes.
For example, for $N=32$ the mixed structure contains $12$ repeating values (out of $20$ distance sequence entries), while the $RS4$ structure contains only $7$ repeating values (out of $16$ entries). In other words, the partial  distance sequence of $RS4$ has better diversity, which may become an advantage in SC decoding. We  return to the codes of Table \ref{tbl:partialDistComb} in our simulation results discussion in Section \ref{sec:simulations}.



\begin{table}
\center
\begin{tabular}{|c|l|l|l|}
   \hline
 $N  $ & \multicolumn{3}{ c| }{Partial Distance Sequences of Polar Code Constructions}\\ \cline{2-4}
   $[bits]$  &$(u+v,v)$  &$RS4$ & $Mixed-RS4$   \\
  \hline
  $4$ & $\left(1,2^{(2)},4\right)$ & $-$ &$\left(1,{\bf 2},4\right)$ \\ \hline
  $8$ & $\left(1,2^{(3)},4^{(3)},8 \right)$&$\left({\bf 1},{\bf 2},{\bf 3},{\bf 4}\right)$ & $\left(1,2,{\bf 2},4,{\bf 4},8 \right)$\\ \hline
  $16$ & $\left(1,2^{(4)},4^{(6)},8^{(4)},16\right)$ & $\left({\bf 1},{\bf 2}^{(2)},{\bf 3}, {\bf 4}^{(2)},{\bf 6},{\bf 8}\right)$ &$\left(1,{\bf 2}^{(2)},4^{(2)},{\bf 4},{\bf 6},{\bf 8}^{(2)},16\right)$ \\

  \hline
  $32$ & $\left(1,2^{(5)},4^{(10)},8^{(10)},16^{(5)},32\right)$ & $\left({\bf 1},{\bf 2}^{(2)},{\bf 3}^{(2)}, {\bf 4}^{(3)},\right.$   & $\left( 1,2,{\bf 2}^{(2)},4^{(2)},{\bf 4}^{(3)},{\bf 6},\right.$ \\
    & &$\left.,{\bf 6}^{(2)},{\bf 8}^{(2)},{\bf 9},{\bf 12}^{(2)},{\bf 16}\right)$  & $\left.,8^{(2)},{\bf 8}^{(3)},12,16,{\bf 16}^{(2)},32\right)$ \\ \hline

    $64$ & $\left(1,2^{(6)},4^{(15)},8^{(20)},16^{(15)},\right.$ & $\left( {\bf 1},{\bf 2}^{(3)},{\bf 3}^{(2)},{\bf 4}^{(5)},{\bf 6}^{(4)},{\bf 8}^{(5)},\right.$& $\left( 1,{\bf{2}}^{(3)},4^{(3)},{\bf 4}^{(3)},{\bf 6}^{(3)},{\bf 8}^{(7)},{\bf 12}^{(2)},\right.$ \\
       &$\left.,32^{(6)},64\right)$ &$\left.,{\bf 12}^{(4)},{\bf 16}^{(3)},{\bf 18},{\bf 24}^{(2)},{\bf 32} \right)$ &$\left.,16^{(3)},{\bf 16}^{(3)},{\bf 18},{\bf 24}^{(3)},{\bf 32}^{(3)}, 64\right)$ \\

  \hline
\end{tabular}
 \normalsize
\caption{Comparison of partial distance sequences induced by different codes based on the binary $(u+v,v)$ construction, the $RS4$ construction (interpreted as binary code) and the mixed-kernels construction of Section \ref{sec:mixKer} with $RS4$ code acting as the auxiliary mapping $g_1(\cdot)$. For $N\in\left\{16,64\right\}$ we included  an additional single layer of the quaternary $(u+v,v)$ inner-code  in the $RS4$ GCC construction. Similarly, for $N\in\left\{8,32\right\}$  we included  a single layer of the binary $(u+v,v)$ inner-code  in the $Mixed-RS4$ construction.  }\label{tbl:partialDistComb}
\end{table}


\subsection{Reduced Decoding Complexity}\label{sec:redDecComplx}
An additional advantage of the mixed-kernels structures may be manifested in terms of the complexity of the SC decoders that operate on them.   We begin by analyzing the SC decoding algorithm time complexity for different polar coding schemes in Subsection \ref{sec:timeComplxty}. In Subsection \ref{sec:spaceComplx} we explore the the memory requirements of these schemes (a.k.a space-complexities). The complexity of SCL decoder implementation is elaborated in Subsection \ref{sec:timeAndSpaceComplextyList}. This discussion justifies our  methodology of fair comparison  between different coding schemes presented in Subsection \ref{sec:fairComparison}. We finally apply this approach when analyzing error-correction performance simulations in Section \ref{sec:simulations}.

\subsubsection{SC Time Complexity} \label{sec:timeComplxty}
The SC decoding steps can be classified into three categories:  \textbf{(SC.a)} likelihood calculations; \textbf{(SC.b)} decision making based on these likelihoods; \textbf{(SC.c)} partial encoding of the decided symbols. The time complexity of  \textbf{(SC.a)} category operations    dominate the time complexity of the entire SC algorithm. This is our justification for regarding the number of operations of \textbf{(SC.a)}  as a good measure of the  SC decoder time complexity.

For a homogeneous kernel of  $\ell$ dimensions over field $F$, the straight-forward calculation of the likelihoods performed on the $i^{th}$ decoding step ($i\in \left[\ell\right]_{-}$) requires $|F|\cdot\left(|F|^{\ell-i-1}-1\right)$ additions and $|F|^{\ell-i}\cdot(\ell-1)$ multiplications (see (\ref{eq:genMxdKernelNonGluedChannel}) where $\theta_i=i+1,\forall i\in\left[\ell\right]_{-}$ for the specification of this naive method). For linear kernels it is possible to perform trellis decoding based on the zero-coset's parity check matrix. In this way the number of additions is $\leq \ell \cdot |F|^{i+1}\cdot \left(|F|-1\right)$ and the number of multiplications is $\leq \ell \cdot |F|^{i+2}$. These bounds do not take into account the fact that many paths in the trellis may be skipped and that some of the nodes in the trellis have input degree $<|F|$. Therefore, the actual number of operations may be reduced significantly, and always be bounded from above by the complexity of the initial naive approach. Table \ref{tbl:complexityOfKernel} summarizes the number of operations for likelihood calculations per each decoding step for different kernels using trellis decoding. Note that due to numerical stability it is preferable to use log-likelihoods instead of likelihoods in the decoding algorithm implementation. In this case the  first number in each tuple in the table should be regarded as the number of log-likelihoods additions. The second item in each tuple is interpreted as the number of $\max^{\star}(\cdot,\cdot)$ operations, where $\max^{\star}\left(\alpha_0,\alpha_1\right)\triangleq\max\{\alpha_0,\alpha_1\} + \log\left(1+\exp\left\{|\alpha_1-\alpha_0|\right\}\right)$.

\begin{table}
\center
\begin{tabular}{|l|c|c|c|c|c|}
   \hline
   Kernel  & \multicolumn{5}{ c| }{Decoding Complexity $\left(\# \cdot,\# +\right)$ }\\ \cline{2-6}
&Step $\#0$  &Step $\#1$ &  Step $\#2$ & Step $\#3$ & Total $\#$ of Operations \\
  \hline
$(u+v,v)$ over $GF(2)$ & $(4,2)$ & $(2,0)$ & - & - & $(6,2)$ \\
$(u+v,v)^{\otimes 2}$ mixed over $GF(2)$ & $(12,6)$ & $(20,4)$ & $(6,0)$ & - & $(38,10)$ \\
$(u+v,v)$ over $GF(4)$  & $(16,12)$ & $(4,0)$ & - & - & $(20,12)$ \\
$G_{RS}(4)$  over $GF(4)$ & $(48,36)$ & $(96,60)$ & $(48,12)$ & $(12,0)$ & $(204,108)$ \\
  \hline
\end{tabular}
 \normalsize
\caption{ Decoding complexity for performing different steps in the kernel likelihood calculations. The leftmost item in each tuple is the number of multiplications and the second item is the number of additions.  The second entry corresponds to the $g_0(\cdot)$ kernel in Section \ref{sec:mixKer}.}\label{tbl:complexityOfKernel}
\end{table}

In order to calculate the total number of operations of the SC algorithm for a code of length $N$ bits, we need to take into account the number of occurrences of each decoding step in the algorithm. This can  be easily achieved by counting the number
of kernels in the code structure of each polar code. Utilizing the  GCC structure of polar codes enables us to easily count using recursion formulae. Specifically, let $a_{(u+v,v)}(N)$ denote the number of occurrences of the $(u+v,v)$ kernel in the $(u+v,v)$ code GCC structure of length $N$ bits. We have for $N=2^n$ bits and $n>1$ that $a_{(u+v,v)}(N) = N/2 + 2\cdot a_{(u+v,v)}(N/2) $ and $a_{(u+v,v)}(2) =1$, therefore $a_{(u+v,v)}(N) = N/2\cdot\log_2(N)$. For the RS4 polar code of length $2\cdot 4^n$ bits for $n>1$ we have $a_{RS4}(N) = N/8+ 4\cdot a_{RS4}(N/4)$, and $a_{RS4}(8)=1$,  as a result $a_{RS4}(N) = N/8\cdot \log_4(N/2)$.

Table \ref{tbl:TotalNumOfOperations} summarizes the number of kernels occurrences   for several types of polar code structures of lengths $N=1024, 2048$ and $4096$ bits. The ${Mixed-RS4}$ option is the construction that was described in Section \ref{sec:mixKer} with $G_{RS}(4)$ as the $g_1(\cdot)$ kernel.  The $(u+v,v)-RS4$ construction means that we have two outer-codes of $RS4$ of length $N/2$ bits that are joined together using one layer of  $(u+v,v)$  over $GF(4)$ inner-code. The $(u+v,v)-Mixed-RS4$ construction means that we have two outer-codes of the ${Mixed-RS4}$ code of length $N/2$ bits that are joined together using one layer of the binary $(u+v,v)$ inner-code. The aforementioned constructions enable us to support different code lengths (this is because the $RS4$ structure is always of length $2\cdot 4^n$ bits and the ${Mixed-RS4}$ structure is always of length $4^n$ bits for some $n>0$). Using the number of occurrences and Table \ref{tbl:complexityOfKernel} we are able to calculate the total number of operations in \textbf{(SC.a)} for each code. Note that here we do not distinguish between additions and multiplications.

\begin{remark}[SC Decoder Shortcuts]
Table \ref{tbl:TotalNumOfOperations} assumes that in SC we have to sequentially  decode all the  elements of the polar code encoder input vector $\bf u$. However, given a code design (i.e. a set of input indices that are frozen), it is possible to reduce the number of decoding operations.  The most obvious "shortcut" is to skip likelihoods calculation of frozen-symbols blocks (since their value is known \textit{a priori}). Small outer-codes of low rates can be decoded as one unit and save calculations (see e.g. \cite[Section D]{Trifonov2012}). Rate $1$ outer-codes can also be decoded efficiently as Almadar-Yazdi and Kschischang suggested \cite{AlamdarYazdi2011}. High-rate linear outer-codes can be efficiently decoded using trellis decoder based on their dual-code (see e.g. Miloslavskaya and Trifonov \cite[Section V]{Miloslavskaya2012}).

We note that application of these shortcuts may affect specific polar code structures differently based on their code design, (see e.g. the comparison between Tables \ref{tbl:complexityListLRate1} and \ref{tbl:complexityListLRate05} in the sequel). Having said that, Table \ref{tbl:TotalNumOfOperations} may still provide the reader with the (crude) time complexity cost differences associated with different polar-code structures.
\end{remark}

 The rightmost column of Table \ref{tbl:TotalNumOfOperations} contains the number of operations for a specific code divided by the number operations for decoding the $(u+v,v)$ code of the same length. By doing so we can quantify the effort in likelihood calculation (and as a result in SC decoding) for each code compared to  Arikan's $(u+v,v)$ scheme. We may observe that the number of operations for SC decoding of the $Mixed-RS4$ code is smaller than the comparable structure of $RS4$ although they have the same polar coding exponent. This is because $g_0(\cdot)$ is much lighter in its SC decoding complexity than the $G_{RS}(4)$ kernel. The second observation that is evident from the table is that Arikan's $(u+v,v)$ structure has significantly lower decoding complexity than both the $Mixed-RS4$ and the $RS4$ based structures. As a consequence, in order to have a fair decoding performance comparison between Arikan's $(u+v,v)$ and the other structures we need to equalize the decoding efforts for these structures. We further discuss this idea  in Section \ref{sec:fairComparison}.

 In this subsection we considered the number of operations of the decoding algorithm as a measure of its complexity. The time it takes to run the decoding algorithm is dependent on both the number of operations and their time duration. This decoding time can be usually reduced by introducing parallelism into the decoding algorithm at the cost of duplicating the processing units and additional control logic. Although SC is a sequential decoding algorithm, most of its decoding steps can be parallelized as indicated \textit{e.g.} by Leroux \textit{et al.} \cite{Leroux2011,Leroux2013}  and the authors \cite[Section 5]{Presman2012}. Therefore the rightmost column in Table \ref{tbl:TotalNumOfOperations} may also indicate a proportional increase in the allocated computation resources (\textit{e.g.} number of logic gates in the hardware implementation) for implementing each of the decoders while keeping the same decoding throughput. This increase also results in a corresponding growth of the decoder power consumption.

 Next we explore the memory requirements (space complexity) of the decoding algorithm.

\begin{table}
\center
\begin{tabular}{|l|l||c|c|c|c||c|c|}
   \hline
 $N$ &  Polar & \multicolumn{4}{ c|| }{$\#$ of  Occurrences } &\multicolumn{2}{ c| }{$\#$ of Operations}\\ \cline{3-8}
[bits] & Code & $(u+v,v)$  &  $(u+v,v)^{\otimes 2}$   & $(u+v,v)$& $G_{RS}(4)$ &  Total & Normalized\\
       &      &    $GF(2)$  & mixed &                         $GF(4)$&   &   &\\
  \hline
    \hline
\multirow{3}*{$1024$} & $(u+v,v)$ & $5120$        & $-$&$-$ & $-$& $40960$ & $1.0$\\
& $(u+v,v)-RS4$ &      $-$   & $-$&$256$ & $512$& $167936$ & $4.1$\\
& $Mixed-RS4$ &      $-$   & $496$& $-$ & $392$& $146112$ & $3.6$\\
  \hline
    \hline
\multirow{3}*{$2048$} & $(u+v,v)$ & $11264$        & $-$&$-$ & $-$& $90112$ & $1.0$\\
& $RS4$ &      $-$  & $-$&$-$ & $1280$& $399360$ & $4.4$\\
& $(u+v,v)-Mixed-RS4$ &    $1024$   & $992$& $-$ & $784$& $300416$ & $3.3$\\
  \hline
   \hline
\multirow{3}*{$4096$} & $(u+v,v)$ & $24576$        & $-$&$-$ & $-$& $196608$ & $1.0$\\
& $(u+v,v)-RS4$ &      $-$   & $-$&$1024$ & $2560$& $831488$ & $4.2$\\
& $Mixed-RS4$ &   $-$ & $2016$& $-$ & $2064$& $740736$ & $3.8$\\
  \hline
\end{tabular}
 \normalsize
\caption{ Number of operations (multiplications and additions) for likelihood calculations in SC decoding for different types of codes. The normalized number of operations is with respect to the binary $(u+v,v)$ code of the same length. }\label{tbl:TotalNumOfOperations}
\end{table}

\subsubsection{SC Space Complexity}\label{sec:spaceComplx}
Table \ref{tbl:memAssets} summarizes the main memory assets required for an efficient time implementation of the SC decoder (as was discussed in the previous subsection). These assets are described for the inner-layer of the GCC structure (see Subsection \ref{subsect:kernelAndCodeDecompose} for the inner-layer definition). In order to derive the total memory size for a scheme of length $N$ symbols we need to add the numbers in the table to the total memory size for the outer-codes of length $N/\ell$ where $\ell$ is the kernel size in symbols. In SC a single outer-code is decoded per GCC layer on each point in time. Therefore we only need to take into account the memory specified for decoding an individual outer-code (i.e. the values in Table \ref{tbl:memAssets} are not to be multiplied by the number of outer-codes per layer).
\begin{table}
\center
\begin{tabular}{|l||l|c|}
   \hline
Memory Name & Description & Size [bits]\\
  \hline
  \textit{LogLikelihoodMem} & Holding the results of (log) likelihood calculation serving as  & $\frac{N\cdot|F|\cdot\lambda }{\log_2(|F|)\cdot\ell}$ \\
     & an input for the next outer-code decoder. &  \\
       \hline
  \textit{PartialEncMem} & Holding   the results of partial   encoding based &  $N$\\
     &on the previous SC decisions. It contains one coset members of & \\
    &  the currently decided sub-code. & \\
  \hline
\end{tabular}
 \normalsize
\caption{SC algorithm memory assets for the \textbf{inner-layer} of the GCC structure for code of length $N$ bits over field $F$ and inner-code size (kernel size) of $\ell$ $F$-symbols. Each log-likelihood is represented by $\lambda$ bits.}\label{tbl:memAssets}
\end{table}

\begin{example}[SC Space Complexity for $RS4$]\label{ex:RS4Mem}
Consider the $RS4$ scheme of length $N = 2\cdot 4^n$ bits. The SC decoder memory size for this structure can be derived by a summation of the numbers in Table \ref{tbl:memAssets} ($|F|=\ell=4$) and the memory size for length $N=2\cdot 4^{n-1}$ bits $RS4$ scheme (as long as $n>1$). Therefore we conclude that the overall memory size is $\frac{N}{2}\cdot \left(\sum_{i=0}^{n-1}\left(\lambda+2\right)4^{-i}\right)=\frac{2}{3} \cdot(\lambda+2)\cdot\left(N-2\right)$ bits, where $\lambda$ is the number of bits assigned for representing a log-likelihood value.
\end{example}

\begin{remark}[Log-Likelihood  vs.  Log-Likelihood Ratio Memory]
In this Section we assume that the channel observations and internal calculations are done in terms of log-likelihoods (LLs), which require   $|F|$ LL  values per $F$-symbol. In SC decoder it is possible to save memory space by subtracting all the LLs by the LL  corresponding to the $0$ element of $F$, and omitting the $LL$ of $0$. These normalized values are called Log-Likelihood Ratios (LLRs). Using LLRs decreases the space required to store   likelihoods by a multiplicative factor of $\frac{|F|}{|F|-1}$, which gives an advantage to the $(u+v,v)$ scheme. However, in order to operate the SCL algorithm (discussed in the next subsection) we have to work with LLs\footnote{ Balatsoukas-Stimming \textit{et al.} \cite{Balatsoukas-Stimming2013a} showed how to use LLRs in  SCL decoding. However, in order to do so each decoding path requires an additional \textit{path-metric} (PM) to be calculated and stored along with the LLRs. Consequently, the number of LLRs and PMs  required to be stored throughout the algorithm is the same as the number of LLs in the standard implementation.}. Since SCL has better performance than SC, we decided to analyze the space complexity using LLs as a preparation for Subsection \ref{sec:timeAndSpaceComplextyList}.
\end{remark}

 For mixed-kernels codes, the outer-codes structures may not be same, therefore the \textit{LogLikelihoodMem} should be taken as the maximum memory required for SC implementation for each of the outer-codes.
\begin{example}[SC Space Complexity for $Mixed-RS4$]\label{ex:MixedKernelMem}
 Consider the \textit{Mixed-RS4} polar code scheme of length $N = 4^n$ bits. In this structure there are two types of outer-codes: the \textit{Mixed-RS4} and the $RS4$ schemes. Therefore, in each layer the memory size should be taken as the maximum memory required for supporting each one of them. Specifically, the mixed-kernel requires \textit{LogLikelihoodMem} of size $\lambda/2\cdot N $ bits for supporting the mixed outer-codes and $\lambda \cdot N$ bits for the $RS4$ outer-code. Therefore, for the inner-layer we need to allocate $\left(\lambda+1\right)\cdot N$ bits. For the complete scheme we take the size of the memory allocated for the inner layer and add to it the maximum size specified for \textit{Mixed-RS4} and \textit{RS4} of length $N/4$ symbols. It can be proven by induction that the \textit{Mixed-RS4} of length of $N/4$ bits requires less memory than the \textit{RS4} code of length $N/4$ quaternary symbols. As a result, the overall scheme employs memory of size $\frac{4}{3}\cdot \left( \lambda+\frac{5}{4}\right)\cdot\left( N-\frac{\lambda+2}{\lambda+\frac{5}{4}}\right)$ bits.
\end{example}
\renewcommand{\arraystretch}{1.3}
\begin{table}
\center
\begin{tabular}{|l|l||l|c|}
   \hline
Polar Code & Restrictions & Memory Size & Normalized Memory Size \\
 &on $N$  [bits] &  [bits] &$\left(N>>1, \lambda = 6 \text{[bits]}\right)$ \\
  \hline
$(u+v,v)$ & power of $2$ & $2\cdot (\lambda + 1)\cdot N$ & $1.00$  \\
\hline
$(u+v,v)-RS4$ &\multirow{2}*{even power of $2$}  & \multirow{2}*{ $\frac{4}{3}\cdot \left( \lambda+\frac{5}{4}\right)\cdot\left( N-\frac{\lambda+2}{\lambda+\frac{5}{4}}\right)$} &  \multirow{2}*{$0.69 $} \\
$Mixed-RS4$& & &\\
\hline
$RS4$ & \multirow{2}*{odd power of $2$}  &   $\frac{2}{3}\cdot\left(\lambda+2\right)\cdot \left(N-2\right)$ & $0.38$ \\
$(u+v,v)-Mixed-RS4$&  & $  \frac{5}{3}\cdot\left(\lambda+\frac{11}{10}\right)\cdot \left(N-\frac{\lambda+2}{\frac{5}{4}\cdot\lambda+\frac{11}{8}}\right) $ & $0.85$ \\
  \hline
\end{tabular}
 \normalsize
\caption{ Total memory requirements for SC decoder of several polar code schemes of length $N$ bits. The log-likelihoods are represented by $\lambda$ bits. For an easy comparison between the schemes, the rightmost column reports the memory size for each scheme divided by the memory size of the $(u+v,v)$ scheme of the same length.}\label{tbl:memTotal}
\end{table}
Table \ref{tbl:memTotal} contains the SC decoders memory consumption  of several polar coding schemes. The  rightmost column of the table contains the quotient of the memory size of each scheme and that of the $(u+v,v)$ scheme for the same (large) code length and $\lambda = 6$ bits representation of the LLs. The table indicates that for  coding schemes of length $N=2\cdot 4^n$ bits the $RS4$ polar code requires  $\sim 38\%$ and $\sim  45\%$ of the memory used by the $(u+v,v)$ and $(u+v,v)-Mixed-RS4$ schemes respectively. For coding schemes of length $N=4^n$ bits, both the \textit{Mixed-RS4}  and  the $(u+v,v)-RS4$ schemes   require $\sim 69\%$ of the memory specified for the $(u+v,v)$ polar code.

\subsubsection{SCL Time and Space Complexities}\label{sec:timeAndSpaceComplextyList}
Tal and Vardy \cite{Tal2011,Tal2012} introduced the SCL decoder that enhances the error-correction performance of the SC algorithm. This performance is improved to greater extent if CRC is concatenated to the polar code.

The SCL algorithm with list of size $L$ considers simultaneously at most $L$ possible prefixes for the transmitted information word (these prefixes are dubbed \textit{decoding-paths}). For each such decoding-path SCL performs the same calculations that are employed in a single SC decoder. In other words, the consumed time and space resources of each decoding step in SC, corresponding to \textbf{(SC.a)} and \textbf{(SC.c)} categories, are grown by at most factor of $L$ in SCL. When the algorithm has to decide on the (non-frozen) information symbol $u_i$ it calculates for each of the surviving decoding-paths $\hat{\bf u}_0^{i-1}$ the likelihood of the prefix when $u_i$ is concatenated to it (i.e. $\hat{\bf u}_0^{i-1}\bullet u_i$, where $u_i \in F$). This step increases the number of paths by a multiplicative factor of $|F|$. The decoder then keeps the  $L$ paths with the highest  likelihood scores.  As a consequence, the complexity of SCL is bounded from above by the addition of two components: \textbf{(i)} $L$ times the complexity of SC   \textbf{(ii)} the total complexity of finding the best $L$ paths (for each of the non-frozen symbols). The complexity of \textbf{(ii)} is negligible compared to the complexity of \textbf{(i)} (assuming that $L$ is fixed, and $N\rightarrow\infty$). Indeed, Tal and Vardy showed that for list size $L$ and Arikan's $(u+v,v)$ code of length $N$ bits the decoding time complexity of the SCL algorithm is $O(L\cdot N\cdot\log N)$ with space complexity of $O(L\cdot N)$. This idea can be further generalized to other homogenous kernels and for the mixed structures (see e.g. \cite{Presman2012}). Observations \ref{observ:SCLTime} and  \ref{observ:SCLSpace} formalize this discussion.


\begin{observation}[SCL Time Complexity]\label{observ:SCLTime}
Let $\mathcal{T}_{SC}$ denote the SC decoding time (measured in number of operations defined in Table \ref{tbl:TotalNumOfOperations}) for a homogenous code of length $N$ bits over field $F$. Let $\mathcal{T}_{SCL}(L)$ denote the decoding time for SCL with list  size $L$ for the same code. We have
\begin{equation}\label{eq:SCLComplexity}
\mathcal{T}_{SCL}(L)\leq L\cdot \mathcal{T}_{SC} +\frac{ N\cdot R }{\log_2(|F|)}\cdot\mu_{T}\left(|F|\cdot L,L \right)+O\left(N\cdot L\cdot \log L  \right),
\end{equation}
where $R$ is the code rate, $\mu_{T}(x,y)$ is the number of operation for finding the $y$ maximal elements in a list of $x$ numbers.
\end{observation}
\proof The first addend on the right hand side is due to the fact that in each point of time there are at most $L$ decoding-paths, each one of them has  time complexity of a single SC decoder. The second addend is due to the selection of best $L$ decoding paths among at most $|F|\cdot L$ candidates. This operation occurs for  non-frozen symbols, and hence it occurs $\frac{ N\cdot R }{\log_2(|F|)}$ times. The third addend in the bound accounts for counters and   pointers handling that occurs in the SCL algorithm. Note that as $N$ grows the second and the third addends in the bound  become negligible compared to the first addend. It
 is known that $\mathcal{T}_{SC} = O(N\cdot \log N)$, assuming that $|F|$ and $\ell$ (the kernel number of dimensions) are constant. The second addend corresponds to an order statics problem and therefore $\mu_{T}\left(|F|\cdot L,L \right) = O(L)$ (see e.g. \cite[Chapter 9]{Cormen2003}). As a conclusion, we may claim that $\mathcal{T}_{SCL} \leq  L\cdot \mathcal{T}_{SC}\cdot \left(1+o(1)\right)$, where $o(1)$   vanishes as $N$ grows. \qed

In Observations \ref{observ:SCLTime} we used an upper-bound to characterize the complexity of SCL in terms of $L$ times the SC complexity. The exact complexity, however, is dependant on the specific code design. In order to understand this remark, note that for each of the first $\left\lceil\log_{|F|}\left(L\right)\right\rceil$ decisions steps on the non-frozen symbols the list size is increased by  a multiplicative factor $ \leq |F|$, from list size $1$ to $L$. Therefore, the number of operations until this decoding point is strictly less than $L$ times the number of operations employed until the same point in the SC algorithm. Tables \ref{tbl:complexityListLRate1} and  \ref{tbl:complexityListLRate05} exemplify this notion.

Table \ref{tbl:complexityListLRate1} summarizes the SCL decoder number of operations (as defined in Table \ref{tbl:TotalNumOfOperations}) for   different rate $R=1$ coding schemes and different list sizes $L$. The number of operations for SC decoding of $(u+v,v)$ (i.e. SCL with list size $L=1$) serves as the normalization reference for the complexity of the other codes having the same length. The leftmost list column ($L=1$) should be recognized as the rightmost column in Table \ref{tbl:TotalNumOfOperations}. It is evident from the table that for fixed $L$, as $N$ increases the number of operations of the SCL decoder tends to be $L$ times the number of operations of the SC algorithm. Note that this is always the case if the index of the non-frozen information symbol number $\left\lceil\log_{|F|}\left(L\right)\right\rceil$   is independent with $N$ (in SCL the number of decoding-paths reaches the maximum list size when decoding this symbol).

Table \ref{tbl:complexityListLRate05} considers several length $N$ bits codes with user information rate of $0.5$ not including $16$ bits of CRC (in other words, the number of unfrozen bits in the information word is $N/2+16$). The codes were designed using Genie-Aided simulations and the number of log-likelihood calculation  operations    was carefully enumerated for each case based on  its design. In this enumeration we assume that operations corresponding to frozen blocks can be skipped, and therefore they are not accounted for. This is the reason why the leftmost list column (describing  $L=1$) is different than the corresponding column in Table \ref{tbl:complexityListLRate1}. It can be seen that indeed the increase in the number of operations is less than $L$ times that of the SC decoder. Furthermore, Tables   \ref{tbl:complexityListLRate1} and  \ref{tbl:complexityListLRate05}  exemplify that the exact time complexity of the SCL algorithm is indeed code design dependant. Notwithstanding, the multiplicative factor $L$ may still serve as a reasonable rule-of-thumb when considering the complexity of SCL compared to SC. Simulations of the error-correction performance of the codes listed in  Table \ref{tbl:complexityListLRate05} are discussed in Section \ref{sec:simulations}.

\begin{remark}[Table \ref{tbl:complexityListLRate1} Revisited]
Table \ref{tbl:complexityListLRate1} example  was given here only for demonstrating the SCL time complexity dependency with the code design (in conjunction with Table \ref{tbl:complexityListLRate05}). The actual results in the table are quite insignificant for the following reasons: \textbf{(i)} All the table's length $N$ bits codebooks are identical and equal to $\{0,1\}^N$. The encoders are different though.
\textbf{(ii)}  The ML decoder for rate $1$ codes is much simpler than the SC decoder \cite{AlamdarYazdi2011}.  Consequently, there is no practical justification for SCL decoding of such codes.
\end{remark}
\begin{table}\center
\begin{tabular}{c l||c|c|c|c|c|c||}
\cline{3-8}\cline{3-8}
     &     & \multicolumn{6}{l||}{Normalized Number of Operations} \\\hline
\multicolumn{1}{||c|}{ \multirow{1}*{$N [bits] $}}& {\backslashbox{Code~~~~~~~~~~~~~~~~~}{$L$~~~~~~~}} & 1       & 2       & 4        & 8        & 16      & 32      \\ \hline\hline
\multicolumn{1}{||c|}{\multirow{3}*{$1024$}}& $(u+v,v)$   & 1.0 & 1.9 & 3.6  & 6.9  & 13.7 & 27.3\\ \cline{2-8}
\multicolumn{1}{||c|}{}&$(u+v,v)-RS4$   & 4.1 & 7.9 & 15.4 & 30.3 & 60.3 & 120.2    \\ \cline{2-8}
\multicolumn{1}{||c|}{}&$Mixed-RS4$ & 3.6 & 7.0 & 13.8 & 27.5 & 54.8 & 109.4    \\ \hline
\hline
\multicolumn{1}{||c|}{\multirow{3}*{$2048$}}& $(u+v,v)$    & 1.0 & 1.9 & 3.6  & 7.0  & 14.0 & 27.8 \\ \cline{2-8}
\multicolumn{1}{||c|}{}&$RS4$ & 4.4 & 8.5 & 16.8 & 33.2 & 66.1 & 131.9\\ \cline{2-8}
\multicolumn{1}{||c|}{}&$(u+v,v)-Mixed-RS4$ & 3.3 & 6.5 & 12.9 & 25.7 & 51.3 & 102.4    \\ \hline
\hline
\multicolumn{1}{||c|}{\multirow{3}*{$4096$}}& $(u+v,v)$    & 1.0 & 1.9 & 3.6  & 7.1  & 14.1 & 28.1  \\ \cline{2-8}
\multicolumn{1}{||c|}{}&$(u+v,v)-RS4$ & 4.2 & 8.2 & 16.0 & 31.8 & 63.3 & 126.3   \\ \cline{2-8}
\multicolumn{1}{||c|}{}&$Mixed-RS4$  & 3.8 & 7.4 & 14.7 & 29.3 & 58.4 & 116.7 \\ \hline
\hline
\end{tabular}
\caption{Normalized number of  SCL  operations  for different list sizes $L$ for different rate $R=1$ coding schemes of length $N$ bits.  For each length $N$ the total number of operations for decoding is divided by the number of operations of $(u+v,v)$ of the same length with $L=1$ (i.e. $SC$). The numbers of operations for $(u+v,v)$ of lengths $N=1024,2048$ and $4096$ bits are $40960, 90112$ and $196608$, respectively.  }
\label{tbl:complexityListLRate1}
\end{table}

\begin{table}\center
\begin{tabular}{c l||c|c|c|c|c|c||}
\cline{3-8}\cline{3-8}
     &     & \multicolumn{6}{l||}{Normalized Number of Operations} \\\hline
\multicolumn{1}{||c|}{ \multirow{1}*{$N [bits] $}}& {\backslashbox{Code~~~~~~~~~~~~~~~~~}{$L$~~~~~~~}} & 1       & 2       & 4        & 8        & 16      & 32      \\ \hline\hline
\multicolumn{1}{||c|}{\multirow{3}*{$1024$}}& $(u+v,v)$    & 1.0     & 1.8     & 3.4      & 6.6      & 12.8    & 25.4    \\ \cline{2-8}
\multicolumn{1}{||c|}{}&$(u+v,v)-RS4$ & 3.6     & 6.6     & 12.6     & 24.7     & 48.8    & 97.0    \\ \cline{2-8}
\multicolumn{1}{||c|}{}&$Mixed-RS4$ & 3.1     & 6.0     & 11.7     & 23.2     & 46.0    & 91.6    \\ \hline
\hline
\multicolumn{1}{||c|}{\multirow{3}*{$2048$}}& $(u+v,v)$    & 1.0 & 1.8 & 3.5  & 6.7  & 13.2 & 26.0  \\ \cline{2-8}
\multicolumn{1}{||c|}{}&$RS4$ & 4.0 & 7.6 & 14.8 & 29.2 & 57.9 & 115.3   \\ \cline{2-8}
\multicolumn{1}{||c|}{}&$(u+v,v)-Mixed-RS4$ & 2.9 & 5.6 & 10.6 & 20.5 & 40.3 & 80.0     \\ \hline
\hline
\multicolumn{1}{||c|}{\multirow{3}*{$4096$}}& $(u+v,v)$     & 1.0 & 1.8 & 3.5  & 6.9  & 13.7 & 26.9    \\ \cline{2-8}
\multicolumn{1}{||c|}{}&$(u+v,v)-RS4$ & 3.7 & 7.0 & 13.5 & 26.5 & 52.5 & 104.5   \\ \cline{2-8}
\multicolumn{1}{||c|}{}&$Mixed-RS4$ & 3.3 & 6.5 & 12.9 & 25.5 & 50.9 & 101.6   \\ \hline
\hline
\end{tabular}
\caption{Normalized number of  SCL  operations  for different list sizes $L$ for different coding schemes of length $N$ bits. The user information rate of the code is $R=0.5$ not including $16$ bits of CRC. For each length $N$ the total number of operations for decoding is divided by the number of operations of $(u+v,v)$ of the same length with $L=1$ (i.e. $SC$). The numbers of operations for $(u+v,v)$ of lengths $N=1024,2048$ and $4096$ bits are $31552, 67976$ and $146656$, respectively. For $N=1024, 2048$ and $4096$ bits the codes were designed using Genie-Aided simulations on BPSK with AWGN at $E_b/N_0=2.5, 2.0$ and $1.8$  dB, respectively.  }
\label{tbl:complexityListLRate05}
\end{table}

\begin{observation}[SCL Space Complexity]\label{observ:SCLSpace}
Let $\mathcal{S}_{SC}$ denote the  SC decoder required memory size   for a homogenous code of length $N$ bits over field $F$. Let $\mathcal{S}_{SCL}(L)$ denote the required memory size for SCL decoder with list size $L$ for the same code. It can be shown that
\begin{equation}\label{eq:SCLSpaceComplexity}
\mathcal{S}_{SCL}(L)\leq L\cdot \mathcal{S}_{SC} +\mu_{S}\left(|F|\cdot L,L \right)+O\left(\log N\cdot L \cdot \log(L)\right),
\end{equation}

where
$\mu_S(x,y)$ is the size of memory used for finding the $y$ maximal elements in a list of $x$ numbers.
\end{observation}
\proof Similarly to Observation \ref{observ:SCLTime}'s proof, the first addend on the right hand side is due to having $L$ decoding paths and the second addend is required for finding the best $L$ decoding paths among at most $|F|\cdot L$ candidates.   The purpose of the third addend is to account for memory holding pointers to the current viable decoding-paths of the SCL algorithm (implementing a tree data-structure, see \cite[Subsection 4.2.2]{Presman2012}).   As $N$ increases  the second and the third addends in the bound  become negligible compared to the first addend. Consequently, we may claim that  $\mathcal{S}_{SCL} (L)\leq  L\cdot \mathcal{S}_{SC}\cdot\left(1+o(1)\right)$, where $o(1)$ vanishes as $N$ grows. \qed

Observation \ref{observ:SCLSpace} also bounds from above the space complexity of SCL decoding as $L$ times the complexity of SC. In this case, for fixed $L$, the memory size cannot be typically reduced significantly by taking into advantage the code design. Therefore, it is reasonable to assume that   $\mathcal{S}_{SCL} (L)\approx  L\cdot \mathcal{S}_{SC}$.

\subsubsection{Fair Comparison and SCL} \label{sec:fairComparison}
Subsections \ref{sec:timeComplxty} and \ref{sec:spaceComplx} demonstrated that different coding schemes have different time and space complexities for the SC decoding algorithm.
Comparing the error-correction performance of SC on these schemes has to take into account the coding system throughput requirement and its memory limitations.   Tables \ref{tbl:TotalNumOfOperations} and  \ref{tbl:complexityListLRate05}  indicate that the $(u+v,v)$ SC decoder has significantly lower time complexity than the other schemes. Consequently given a time-complexity constraint, it is reasonable to utilize a  more enhanced decoding scheme (having higher time complexity still meeting the constraint) for the $(u+v,v)$ code.  One possibility for accomplishing this idea  is by increasing the length of the $(u+v,v)$ code. Since the error-correction performance of decoding algorithms typically improves as longer codes are employed, this technique may be useful for surpassing the original decoder's performance.    However, this approach is problematic because in many cases the code length is a requirement of the communication system and cannot be increased\footnote{Let us consider two simple scenarios exemplifying the dependency of the code length with other features of the communication system: \textbf{(i)} The code length determines the number of bits required for transmitting a single bit over the channel. Accordingly it influences the transmission latency of the communication system. \textbf{(ii)} In storage applications, the code length defines the minimum size of information that needs to be retrieved from the device for reliably fulfilling a  user's read request. Consequently it affects the system read latency.}.
Therefore, in this correspondence we use the following comparison guidelines: \textbf{(i)} All the compared coding schemes  will  have an equal code length. This length is understood to be the maximum value still complying with the communication system specifications.  \textbf{(ii)}  In order to equalize the decoding complexities of different schemes we  employ  SCL with different list sizes.

The discussion in Section \ref{sec:timeAndSpaceComplextyList} has set the stage for performing fair comparisons by applying guideline  \textbf{(ii)}.    Let $\mathcal{C}_1$ and $\mathcal{C}_2$ be two coding schemes of equal length decoded by the SCL algorithm with list sizes $L_1$ and $L_2$, respectively. We may consider two extreme case studies.
\begin{itemize}
  \item \textbf{Case Study I (CS-I):}  The decoder implementation is limited by the required throughput or by the number of gates of its hardware implementation\footnote{It is assumed that the throughput specification may be accomplished by introducing sufficient level of decoding parallelism, see Subsection \ref{sec:timeComplxty}.}. In this scenario we should choose $L_1$ and $L_2$ such that their time complexities (and thereby their implication on the computation resources or logical gates count) will meet the requirements. For this task, an analysis such as the one depicted in Table \ref{tbl:complexityListLRate05} may be regarded as a useful reference.
  \item \textbf{Case Study II (CS-II):} The implementation is limited by the algorithm memory size. In this scenario we may use Table \ref{tbl:memTotal} and Observation \ref{observ:SCLSpace} for choosing $L_1$ and $L_2$ such that memory requirements are  met.
\end{itemize}

   The \textbf{(CS-I)} and \textbf{(CS-II)} scenarios illustrate two extremal limitations. Typically, a system designer may have a requirement on the throughput while experiencing limitations on the  computation resources/logical-gates count and the allowed memory size. Hence  his challenge is to choose the solution that meets these specifications and demonstrate the best performance according to some criteria\footnote{ Typical optimization criteria may comprise the ones considered in this section: maximum throughput, minimum logical gate count and minimum memory size.  Additional criteria may also include e.g. maximum error-correction performance and minimum  power consumption.}. In such scenarios, considering the possible solutions when only one of the specifications is taken into account (i.e. reducing the problem to \textbf{(CS-I)} or \textbf{(CS-II)})  and then selecting only the configurations that satisfies also the other constraints will give the set of valid designs from which the best option is to be picked.

In the sequel we present simulation results of the schemes from Table \ref{tbl:complexityListLRate05} and use the above comparison guidelines to demonstrate that the $Mixed-RS4$ structures outperform Arikan's $(u+v,v)$ codes in SCL decoding.

\section{Simulation Results}\label{sec:simulations}

Proposition \ref{propo:rateForMxdKernelGeneral} implies  that when considering
the exponent as a measure of the polarization rate, the behavior of a mixed-kernels structure is the same as the behavior of the weakest kernel from its surviving kernels.
However, the exponent is an asymptotic measure and it may fail
capturing the performance of a polar coding scheme for a finite block length $N$. Indeed, Section \ref{sec:mixedPotAdvantage} suggests that employing mixed-kernels may lead to improved error-correction performance due to a better code decomposition, with moderate SCL decoding complexity.

In this section we demonstrate this performance improvement conjecture  for rate $0.5$ codes of block length $N=1024, 2048$ and $4096$ bits that were listed in Table \ref{tbl:TotalNumOfOperations}. The codes were simulated over the AWGN channel with BPSK modulation. The design of all the codes was done by Genie-Aided (GA) simulations performed on one of the SNR points of each simulation. We used SCL with different list sizes (indicated by the parameter $L$), and different outer CRC codes. We tried both CRC codes of $8$ bits and of $16$ bits and present in the figures, the CRC that gave the best results. 
For each simulation point we collected at least $100$ frame error events.

\begin{figure}
\centering
\begin{subfigure}{.5\textwidth}
  \centering
  \includegraphics[scale = 0.37]{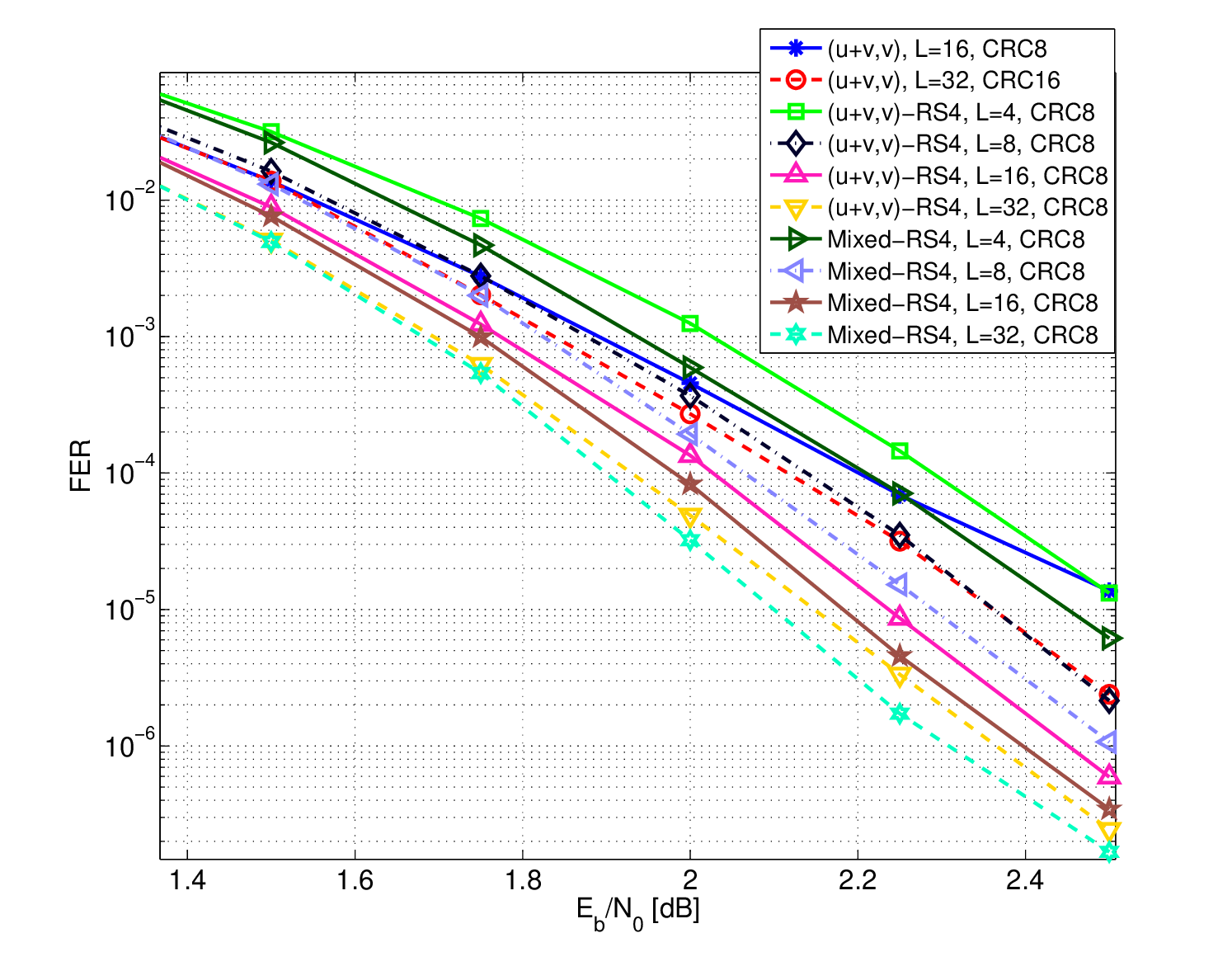}
  \caption{$N=1024, \left(E_b/N_0\right)_{design} = 2.25 [dB]$}
  \label{fig:simResN1024R05}
\end{subfigure}%
\begin{subfigure}{.5\textwidth}
  \centering
  \includegraphics[scale = 0.37]{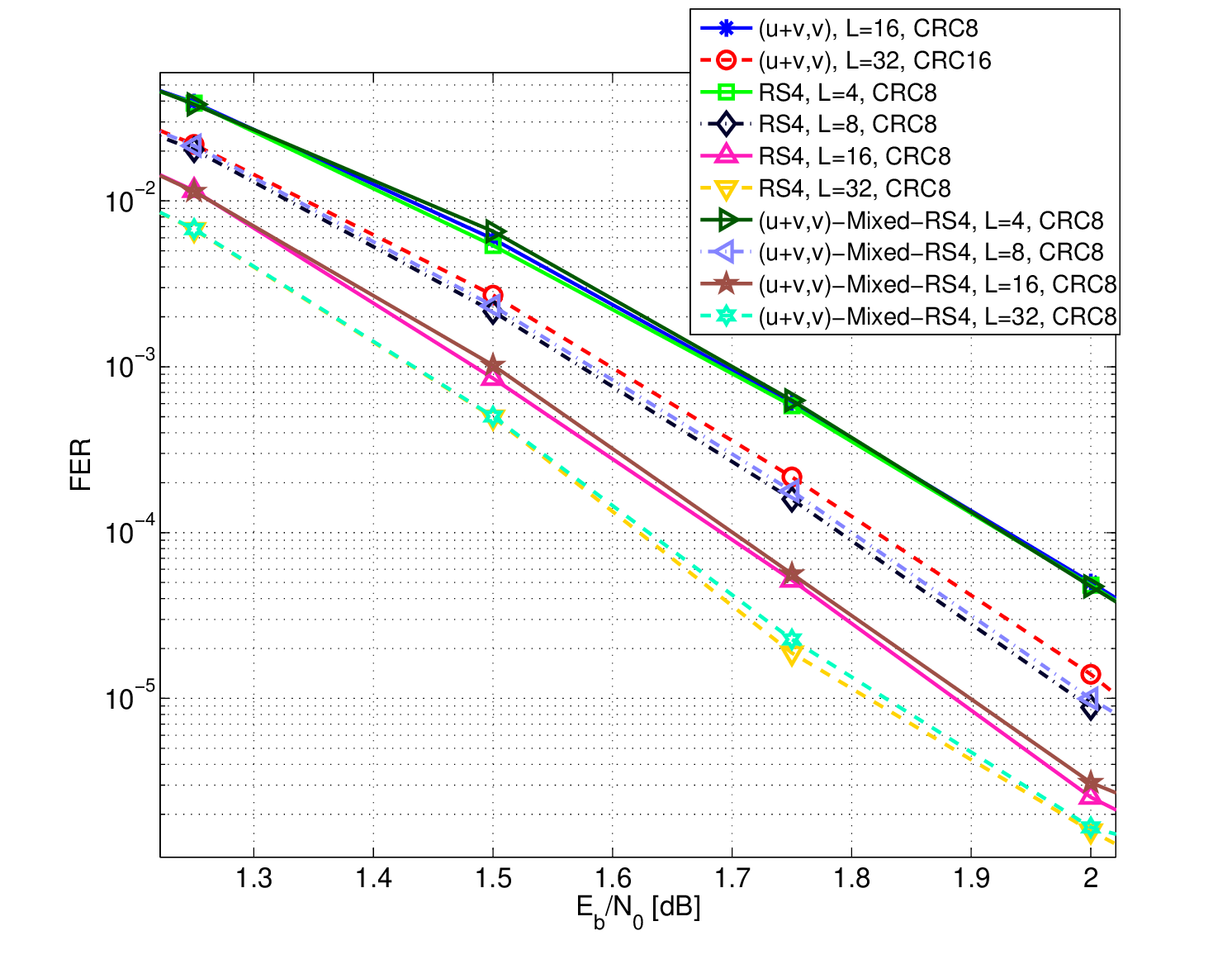}
  \caption{$N=2048, \left(E_b/N_0\right)_{design} = 2.00 [dB]$}
  \label{fig:simResN2048R05}
\end{subfigure}
\begin{subfigure}{.5\textwidth}
  \centering
  \includegraphics[scale = 0.37]{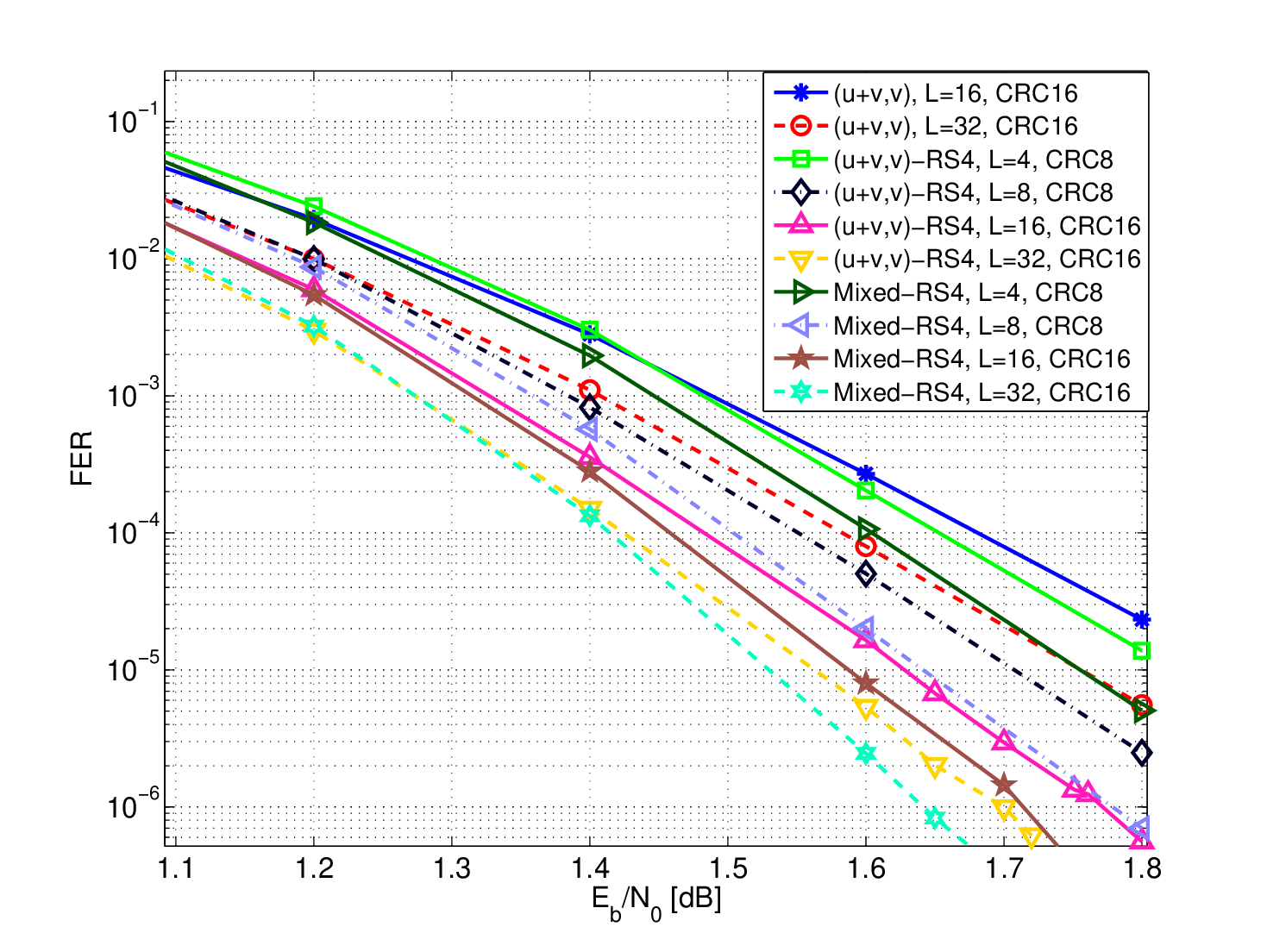}
  \caption{$N=4096, \left(E_b/N_0\right)_{design} =1.80 [dB]$}
  \label{fig:simResN4096R05}
\end{subfigure}%
\begin{subfigure}{.5\textwidth}
  \centering
  \includegraphics[scale = 0.37]{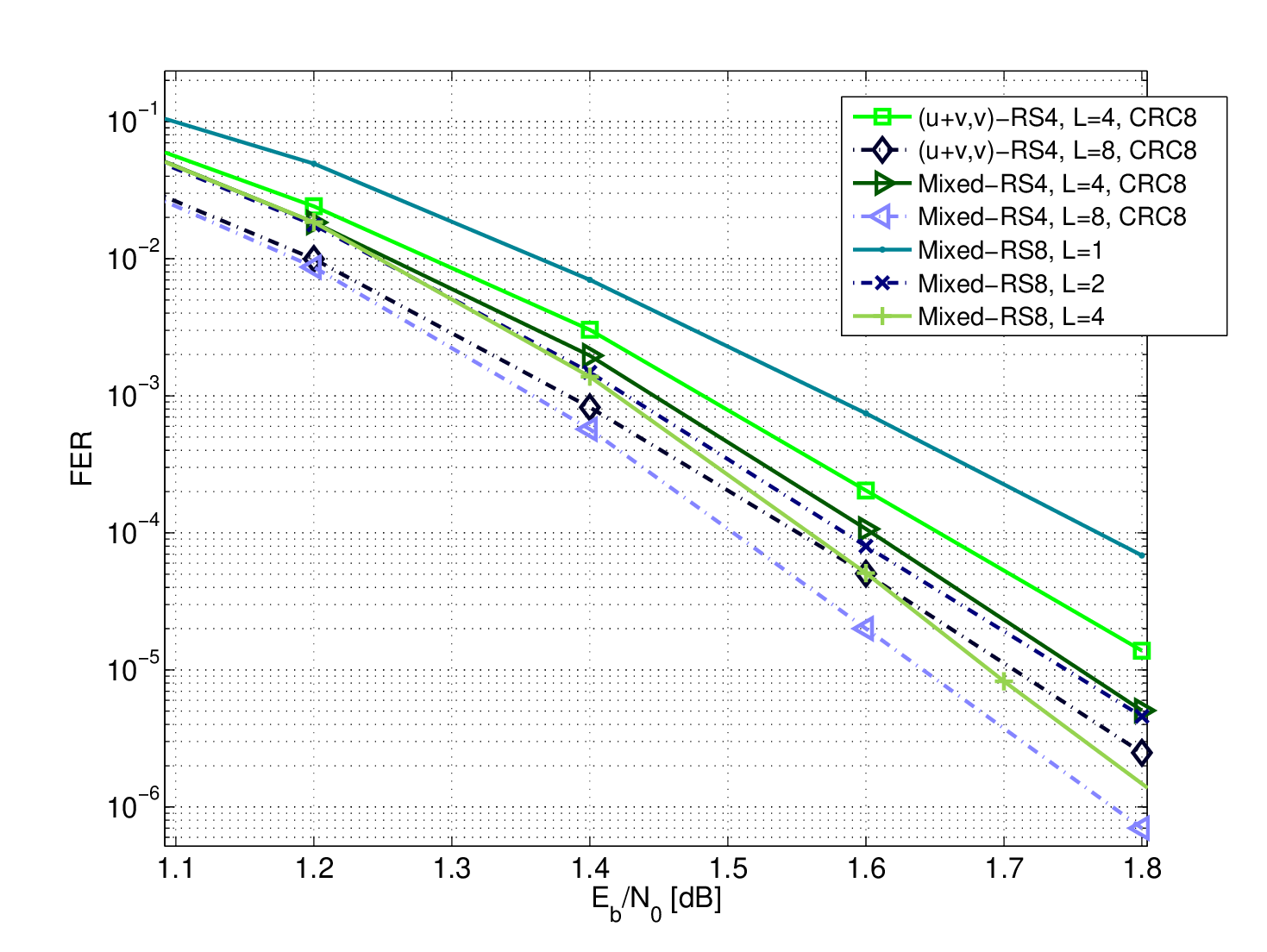}
  \caption{$N=4096$, including $Mixed-RS8$ with $\left(E_b/N_0\right)_{design} = 1.75 [dB]$}
  \label{fig:simResN4096R05RS8}
\end{subfigure}
\caption{FER simulation of the SCL decoding algorithms of various rate $0.5$ codes  of different lengths $N$ bits from Table \ref{tbl:TotalNumOfOperations} over AWGN with BPSK modulation.  Each code was designed by running GA simulation at SNR point $\left(E_b/N_0\right)_{design}$.  The list size is indicated by $L$ and the CRC length is indicated by the number following the CRC label (i.e. CRC8 and CRC16 for $8$ and $16$ bits CRCs, respectively).}\label{fig:FERSimResults}
\end{figure}

Figure \ref{fig:simResN1024R05} depicts the frame-error-rate (FER) results simulated for $N=1024$ bits codes.   The $(u+v,v)$  list sizes   were of $16$ and $32$.   We consider the two case-studies from  Subsection \ref{sec:fairComparison}. \textbf{(CS-I)}:  using Table \ref{tbl:complexityListLRate05} we may deduce that the $(u+v,v)$  list sizes of  $16$ and $32$  should be compared with the list sizes of $4$ and $8$ respectively of the other schemes. It is evident that the ${Mixed-RS4}$ achieves better  error-correction performance than the $(u+v,v)-RS4$ alternative with smaller complexity.  On the high SNR points the $(u+v,v)$  achieves similar error-correction performance to the ${Mixed-RS4}$ with an apparent trend that the   ${Mixed-RS4}$ outperforms the $(u+v,v)$ as the SNR increases.
\textbf{(CS-II)}: using Table \ref{tbl:memTotal} we can compare the schemes with the same list sizes. Here there is a clear advantage of the mixed schemes compared to their corresponding candidates from the other schemes. We note that this is achieved with $\approx 40\%$ less memory resources compared to the $(u+v,v)$ polar code.

The trend that was illustrated in the last paragraph is enhanced  in Figure \ref{fig:simResN4096R05} that depicts the FER results for $N=4096$ bits codes. Indeed, the $Mixed-RS4$ codes outperform their other comparable coding schemes both according to  \textbf{(CS-I)} and \textbf{(CS-II)}. Figure \ref{fig:simResN4096R05RS8} contains simulation results of the $Mixed-RS8$ construction that was discussed in Example \ref{ex:l8mixedExample}.
Note that  Although the $Mixed-RS8$ space complexity is lower than that of the other schemes, its time complexity is much higher.
Consequently, considering \textbf{(CS-I)} scenarios, the figure suggests that $Mixed-RS4$ constructions are preferable.


Figure \ref{fig:simResN2048R05} depicts the FER results simulated for $N=2048$ bits codes. The $(u+v,v)$ curve of SCL with $L=32$ was also simulated by Tal and Vardy \cite[Figure 1]{Tal2012}. Comparing the  $(u+v,v)-Mixed-RS4$ scheme with list sizes of $4$ and $8$ with $(u+v,v)$  with list sizes of  $16$ and $32$  respectively indicates that both of them have similar FER while  the first implementation requires less complexity compared to the second one. The $RS4$ scheme has similar FER results to that of $(u+v,v)-Mixed-RS4$ using the same list size. The $(u+v,v)-Mixed-RS4$  scheme has smaller SCL time complexity than $RS4$ with the same list size $L$ (reduction of $> 29 \%$ in the number of operations, according to Table \ref{tbl:complexityListLRate05}). On the other hand, the $RS4$ scheme has less SCL space complexity compared to the $(u+v,v)-Mixed-RS4$ with same list size $L$ (reduction of $\sim 55\%$ of the memory size, according to Table \ref{tbl:memTotal}).


\section{Summary and Conclusions}\label{sec:sumAndConcold}
Mixed-kernels constructions of polar codes were introduced and analyzed in this paper. We began by providing conditions for polarization of the mixed-kernels structures based on their constituent kernels. Then we turned to calculate their polar coding exponent. Both the polarization property and the rate of polarization are asymptotic  in the code length. Considering finite length instances of these codes suggests possible advantages in the error-correction performances and the decoder complexity.

Throughout the paper we used an example based on $(u+v,v)^{\otimes 2}$ kernel and a quaternary kernel of size $4$. Our preliminary intention in using this example was to simplify the introduction of mixed-kernels and their relevant notations and definitions. Simulations of the SCL decoding algorithm of this example (taking $G_{RS}(4)$ as the quaternary kernel) indicate that this scheme is attractive both in terms of error-correction performance and in terms of decoder complexity. Indeed, in many cases this $Mixed-RS4$ codes demonstrate better error-correction/complexity tradeoff than the known polar code schemes.
\bibliographystyle{IEEEtran}

\section*{Acknowledgements}\label{sec:ack}
The authors would like to thank Dr. Nissim Halabi for helpful discussions and for his contribution to the development of the simulation software that produced Section \ref{sec:simulations} results.

\bibliography{IEEEabrv,bibTexPolar}

\normalsize

\end{document}